\documentclass[12pt]{article}
\usepackage{amsmath,amssymb,exscale,epsf}  

\def \gappeq{\mathrel{\rlap {\raise.5ex\hbox{$>$}}{\lower.5ex
\hbox{$\sim$}}}}
\def \permil{\%\raise.20ex \hbox{$_0$}
\def\lappeq{\mathrel{\rlap{\raise.5ex\hbox{$<$}}
\lower.5ex\hbox{$\sim$}}}}	
	
\begin{document}	
\topmargin -1.0cm
\oddsidemargin -0.8cm
\evensidemargin -0.8cm
\pagestyle{empty}
\begin{flushright}
UNIL-IPT-02-06\\ 
\end{flushright}
\vspace*{5mm}

\begin{center}

{\Large\bf Localizing gravity on a 't Hooft-Polyakov monopole 
in seven dimensions}\\

\vspace{1.0cm}

{\large Ewald Roessl\footnote{Email: ewald.roessl@ipt.unil.ch}
and Mikhail Shaposhnikov
\footnote{Email: mikhail.shaposhnikov@ipt.unil.ch}}\\
\vspace{.6cm}
{\it {Institute of Theoretical Physics\\ University of Lausanne\\ 
CH-1015 Lausanne, Switzerland}}
\vspace{.4cm}
\end{center}

\vspace{1cm}  
\begin{abstract} 
 
We present regular solutions for a brane world scenario in the form
of a 't Hooft-Polyakov  monopole living in the three-dimensional
spherical symmetric transverse space of a seven-dimensional 
spacetime. In contrast to the cases of a domain-wall in five
dimensions and a string in six dimensions, there  exist
gravity-localizing solutions for both signs of the bulk cosmological
constant. A detailed  discussion of the parameter space that leads to
localization of gravity is given. A point-like monopole limit is
discussed. 

\end{abstract}

\vfill

\eject
\pagestyle{empty}
\setcounter{page}{1}
\setcounter{footnote}{0}
\pagestyle{plain}


\section{Introduction} 

Recently, there has been renewed interest in brane-models in which
our world is represented as a $3+1$-dimensional submanifold  (a
3-brane) living in a higher-dimensional space-time \cite{RS2,akama}.
This idea provides an alternative to Kaluza-Klein compactification
\cite{KK} and gives new insights to a construction of low energy 
effective theory of the fields of the standard model \cite{RS2,DS} 
and gravity
\cite{Arkani-Hamed:1998rs,rusu2}. Moreover, it
may shed light on gauge hierarchy problem
\cite{Arkani-Hamed:1998rs,rusu1} and on cosmological constant problem
\cite{Rubakov:1983bz}-\cite{Dvali:2002pe}.

In string theory Standard Model fields are localized on D-branes -
\cite{branes}, whereas from the point of view of field theory brane
model could be realized as a topological defect formed by scalar and
gauge fields being a solution to the classical equations of motion of 
the coupled Einstein -- Yang-Mills -- scalar field equations. In the
latter case one should be able to construct a solution leading to a
regular geometry and localizing fields of different spins, including
gravity.

Quite a number of explicit solutions is already known. In
five-dimensions a real scalar field forming a domain wall may serve
as a model of 3-brane \cite{DeWolfe:1999cp}. Higher dimension
topological defects can be qualitatively different from a
five-dimensional case from the point of view of localization of
different fields on a brane. So, strings in six space-time
dimensions were considered in \cite{ck}-\cite{Harvey}. In particular,
solutions corresponding to a thin local string together  with
fine-tuning relations (similar to the Randall-Sundrum domain-wall 
case) were found in \cite{gs} (see also \cite{KP}). A numerical
realization confirming the general results of \cite{gs} in a
singularity-free geometry and for the case of the Abelian Higgs model
has been worked out in \cite{Harvey}. Moving to even higher
dimensions may be of interest because of a richer content of
fermionic zero modes and because of more complicated structure of
transverse space. In the framework of KK compactification monopoles
in seven dimensions  and instantons in eight dimensions were
discussed in \cite{Randjbar-Daemi:1982hi,Randjbar-Daemi:1983qa}, and
brane-world scenarios in higher dimensions in \cite{ov},
\cite{GRS}-\cite{Randjbar-Daemi:2000cr}.

In \cite{GRS} we considered a general point-like spherically
symmetric topological defect as a model of 3-brane and formulated
conditions that are necessary for gravity localization on it. A
transition from a regular solution to the classical equations of
motion to a point-like limit is in fact quite non-trivial for six and
higher dimensions (see a detailed discussion for a string case in
\cite{Harvey,KP}). The aim of the present paper is to provide an
existence proof of a possibility of gravity localization on a
regular three-dimensional defect -- 't~Hooft-Polyakov monopole in
seven dimensions, to study the parameter-space of a model that leads
to gravity localization and to formulate exactly the meaning of
point-like monopole limit. We confirm entirely the previous results,
in particular, a possibility of gravity localization on a monopole
embedded in a space with both signs of a bulk cosmological constant.

The paper is organized as follows. In section \ref{EQFQ} we present
the $SO(3)$ invariant Georgi-Glashow model (having 't Hooft-Polyakov
monopoles as flat spacetime solutions \cite{tHooft},\cite{Polyakov})
coupled to gravity in seven dimensions. The Einstein equations and
the field equations are obtained in the case of a generalized  
't~Hooft-Polyakov ansatz for gauge and scalar fields.   Boundary
conditions are discussed in section \ref{BC}, the asymptotic behavior
of the solutions at the origin in section  \ref{BSO}, at infinity in
section \ref{BSI}.  In section \ref{FTR} we give the relations
between the brane tension components necessary for warped
compactification. Section \ref{N} presents numerical results
omitting  all technical details. We give explicit sample solutions
in  section \ref{samplesol} and  discuss their general dependence on
the parameters of the model. We then present the fine-tuning surface
(the  relation between the  independent parameters of the model
necessary for gravity  localization) in section \ref{NFTS}. Sections
\ref{PSL} and \ref{FML} treat the Prasad-Sommerfield limit and the
point-like monopole limit,  respectively. While in the former case no
gravity localizing solutions exist, in the latter case we demonstrate
a possibility of the choice of the model parameters that leads to a
fundamental Planck scale in TeV range and small modifications of the
Newton's law, while well within the range of applicability of
classical gravity. We conclude in section \ref{C}. In Appendix~A a
derivation of the fine-tuning relations is given, whereas in Appendix
B a discussion of the numerical details can be found. 

\section{Field equations} \label{EQFQ}
The action for the setup considered in this paper is a
straightforward generalization of a gravitating 't Hooft-Polyakov
monopole in 4 dimensions (which has been extensively studied in the
past \cite{nwp}-\cite{bfm}) to the case  of 
seven-dimensional spacetime:
\begin{equation} 
\label{action}
  S=S_{\mbox{\tiny\it gravity}}+S_{\mbox{\tiny\it brane}} .
\end{equation}
Here $S_{\mbox{\tiny\it gravity}}$ is the seven-dimensional
Einstein-Hilbert action:
\begin{equation} 
\label{act_grav}
S_{\mbox{\tiny\it gravity}}=\frac{M_7^5}{2} \int d^7 x \sqrt{-g}
\left( R - \frac{2 \Lambda_7}{M_7^5} \right)~,
\end{equation}
$g$ is the determinant of the metric $g_{M N}$ with signature
$(-++++++)$.
We use the sign conventions for the Riemann tensor of  \cite{wigner}.
Upper case latin indices $M,N$ run over $0 \ldots 6$, lower case
latin indices $m,n$ over $4 \ldots 6$ and greek indices $\mu, \nu$
over $0 \ldots 3$. The parameter $M_7$ denotes the fundamental
gravity scale, $\Lambda_7$ is the bulk cosmological constant and
$S_{\mbox{\tiny\it brane}}$ is the action of Georgi and Glashow 
\cite{GG} containing SU(2) gauge field $W_M^{\tilde{a}}$ and a scalar
triplet  $\Phi^{\tilde{a}}$ (we denote  group indices by $\tilde{a},
\tilde{b}, \tilde{c} =1 \ldots 3$) :
\begin{equation} 
\label{act_brane}
 S_{\mbox{\tiny\it brane}}=\int d^7 x \sqrt{-g} \, \mathcal{L}_m
\quad \mbox{with} \quad \mathcal{L}_m= -\frac{1}{4} G_{M
N}^{\tilde{a}} G^{\tilde{a} M N}-\frac{1}{2} \mathcal{D}_M
\Phi^{\tilde{a}} \mathcal{D}^M \Phi^{\tilde{a}}-\frac{\lambda}{4}
\left( \Phi^{\tilde{a}} \Phi^{\tilde{a}} -\eta^2 \right)^2~,
\end{equation}
where $\eta$ is the vacuum expectation value of the scalar field and 
$\mathcal{D}_M$ is a covariant derivative,
\begin{equation} 
\label{cov_der} 
\mathcal{D}_M \Phi^{\tilde{a}}=\partial_M \Phi^{\tilde{a}} + e \,
\epsilon^{\tilde{a}\tilde{b}\tilde{c}}\, W_M^{\tilde{b}}
\Phi^{\tilde{c}}. 
\end{equation}
Furthermore one has
\begin{equation} 
\label{GaugeField}
  G_{M N}^{\tilde{a}}=\partial_M W_N^{\tilde{a}}-\partial_N
  W_M^{\tilde{a}}+e \, \epsilon^{\tilde{a}\tilde{b}\tilde{c}}  \,
  W_M^{\tilde{b}} W_N^{\tilde{c}}.
\end{equation}
The $SO(3)$ symmetry is spontaneously broken down to $U(1)$. The
monopole corresponds to the simplest topologically nontrivial field
configuration with unit winding number. The Higgs mass is given by
$m_H=\eta \, \sqrt{2 \lambda}$. Two of the gauge fields acquire a
mass $m_W=e \, \eta$.

The general coupled system of Einsteins equations and the equations
of motion for the scalar field and the gauge field following from the
above action are
\begin{eqnarray}  
  R_{M N} -\frac{1}{2} g_{M N} R + \frac{\Lambda_7}{M_7^5} g_{M N}
  &=& \frac{1}{M_7^5} T_{M N}\, ,\label{Einstein} \\
  \frac{1}{\sqrt{-g}} \mathcal{D}_M \left( \sqrt{-g} \, \mathcal{D}^M
  \Phi^{\tilde{a}} \right) &=& \lambda \, \Phi^{\tilde{a}}  \left(
  \Phi^{\tilde{b}} \Phi^{\tilde{b}}-\eta^2\right) \, ,
  \label{EoMscalar} \\ \frac{1}{\sqrt{-g}} \mathcal{D}_M \left(
  \sqrt{-g} \, G^{\tilde{a} M N} \right) &=& -e \,
  \epsilon^{\tilde{a}\tilde{b}\tilde{c}} \left( \mathcal{D}^N
  \Phi^{\tilde{b}} \right) \Phi^{\tilde{c}} \, , 
  \label{EoMgauge} 
\end{eqnarray}
where the stress-energy tensor $T_{M N}$ is given by
\begin{equation} 
\label{EnMom}
  T_{MN}= -\frac{2}{\sqrt{-g}} \frac{\delta S_{\mbox{\tiny\it
brane}}}{\delta g^{M N}}= G_{M L}^{\tilde{a}} G_{N}^{\tilde{a}\,L}+
\mathcal{D}_M \Phi^{\tilde{a}} \mathcal{D}_N \Phi^{\tilde{a}} + g_{M
N} \mathcal{L}_m \, .
\end{equation}

We are interested in static monopole-like solutions to the set of
equations (\ref{Einstein})-(\ref{EoMgauge}) respecting both, 
$4D$-Poincar\'e invariance on the brane and rotational invariance in
the transverse space. The fields $\Phi^{\tilde{a}}$ and
$W_M^{\tilde{a}}$ (and as a result $G_{M N}^{\tilde{a}}$) should not
depend on coordinates on the brane $x^{\mu}$. The brane is
supposed to be located at the center of the magnetic monopole. A
general non-factorisable ansatz for the metric satisfying the above
conditions is 
\begin{equation} 
\label{metric}
  ds^2=M^2(\rho) g^{(4)}_{\mu\nu} dx^{\mu} dx^{\nu} + d\rho^2+
L^2(\rho) \left( d\theta^2+\sin^2 \theta \, d\varphi^2 \right) \, ,
\end{equation}
where $g^{(4)}_{\mu\nu}$ is the four-dimensional metric that 
satisfies the $4D$ Einstein equations with an arbitrary cosmological
constant $\Lambda_{\mbox{\small{phys}}}$ \cite{Rubakov:1983bz}.  In
this paper we will only consider the case of
$\Lambda_{\mbox{\small{phys}}}=0$ and we take $g^{(4)}_{\mu\nu}$  to
be the Minkowski metric $\eta_{\mu \nu}$ with signature $(-+++)$. 

With the use of spherical coordinates for transverse space
\begin{eqnarray} 
\label{intvec}
  \vec{e}^{\,\,\rho}&=&\left( \sin\theta \cos\varphi,\sin\theta
  \sin\varphi,\cos\theta \right) \, , \\
  \vec{e}^{\,\,\theta}&=&\left( \cos\theta \cos\varphi,\cos\theta
  \sin\varphi,-\sin\theta \right)\, , \\
  \vec{e}^{\,\,\varphi}&=&\left( -\sin\varphi,\cos\varphi,0 \right)\,,
\end{eqnarray}
the 't Hooft-Polyakov ansatz is :
\begin{equation} 
\label{tHooftPolysphere}
  \vec{W}_{\rho}=0, \quad 
  \vec{W}_{\theta}=-\frac{1-K(\rho)}{e}\,\vec{e}^{\,\,\varphi},
  \quad   \vec{W}_{\varphi}=\frac{1-K(\rho)}{e}\,\sin\theta \,
  \vec{e}^{\, \, \theta}, \quad \vec{W}_{\mu}=0 \quad
  \vec{\Phi}=\frac{H(\rho)}{e \rho} \vec{e}^{\, \, \rho} \, ,
\end{equation}
with flashes indicating vectors in internal $SO(3)$ space.

Using the ansatz (\ref{tHooftPolysphere}) for the fields together
with the metric (\ref{metric}) in the coupled system of differential
equations (\ref{Einstein})-(\ref{EoMgauge}) gives
\begin{eqnarray}
  3\, \frac{M''}{M}+ 2\, \frac{\mathcal{L}''}{\mathcal{L}}+ 6\,
  \frac{M'\,\mathcal{L}'}{M\,\mathcal{L}} +  3\, \frac{{M'}^2}{M^2} +
  \frac{{\mathcal{L}'}^2}{{\mathcal{L}}^2} -\frac{1}{\mathcal{L}^2}
  &=&  \beta \left(\epsilon_0-\gamma \right) \, ,
  \label{EinsteinP0}\\ 8\, \frac{M'\,\mathcal{L}'}{M\,\mathcal{L}} +
  6\, \frac{{M'}^2}{M^2} +  \frac{{\mathcal{L}'}^2}{{\mathcal{L}}^2}
  - \frac{1}{\mathcal{L}^2} &=&  \beta \left(\epsilon_\rho-\gamma
  \right) \, , \label{EinsteinPrho}\\  4\, \frac{M''}{M}+
  \frac{\mathcal{L}''}{\mathcal{L}}+ 4\,
  \frac{M'\,\mathcal{L}'}{M\,\mathcal{L}} +  6\, \frac{{M'}^2}{M^2}
  &=&  \beta \left(\epsilon_\theta-\gamma \right) \, ,
   \label{EinsteinPtheta}
\end{eqnarray}
\begin{eqnarray} 
J''+2\left(2\, \frac{M'}{M}+\frac{\mathcal{L}'}{\mathcal{L}} \right)
  J'-2\,\frac{K^2 J}{\mathcal{L}^2}&=& \alpha J \left(J^2-1\right) \,
  ,\label{EqMovphi}\\ K''+\frac{K \left( 1- K^2
  \right)}{\mathcal{L}^2}+4\, \frac{M'}{M} K' &=& J^2 K \,
  .\label{EqMovW}
\end{eqnarray}
where primes denote derivatives with respect to the transverse radial
 coordinate $r$, rescaled by the mass of the gauge boson $m_W$:
\begin{equation}
  r=m_W \rho = \eta \, e \, \rho.
\end{equation}
All quantities appearing in the above equations are dimensionless,
including $\alpha, \beta$ and $\gamma$:
\begin{equation}
  \alpha = \frac{\lambda}{e^2}=\frac{1}{2}
  \left(\frac{m_H}{m_W}\right)^2 \, , \quad \beta =
  \frac{\eta^2}{M_7^5} \, , \quad \gamma = \frac{\Lambda_7}{e^2
  \eta^4} \, , \quad \epsilon_i = \frac{f_i}{e^2 \eta^4} \, ,
\end{equation}
with 
\begin{equation} 
\label{f_def}
  T^{\mu}_{\nu}= \delta^{\mu}_{\nu} f_0 \,, \quad
  T^{\rho}_{\rho}= f_\rho \, , \quad 
  T^{\theta}_{\theta}= f_\theta \,, \quad
  T^{\varphi}_{\varphi}= f_\varphi \,, \quad 
  \mathcal{L}=L m_W \, ,\quad  
  J=\frac{H}{\eta \, e \, \rho}.
\end{equation}
The dimensionless diagonal elements of the stress-energy tensor are
given by
\begin{eqnarray} \label{enmomelements1}
  \epsilon_0&=&-\left[
  \frac{{K'}^2}{\mathcal{L}^2}+\frac{\left(1-K^2\right)^2}{2
  \mathcal{L}^4}+\frac{1}{2} {J'}^2 +  \frac{J^2
  K^2}{\mathcal{L}^2}+\frac{\alpha}{4} \left( J^2-1\right)^2 
  \right]\, , \\ \label{enmomelements2} \epsilon_\rho&=&
  \frac{{K'}^2}{\mathcal{L}^2}-\frac{\left(1-K^2\right)^2}{2
  \mathcal{L}^4}+\frac{1}{2} {J'}^2 -  \frac{J^2
  K^2}{\mathcal{L}^2}-\frac{\alpha}{4} \left( J^2-1\right)^2 \, ,
  \\
  \label{enmomelements3} 
  \epsilon_\theta&=&
  \frac{\left(1-K^2\right)^2}{2 \mathcal{L}^4}-\frac{1}{2} {J'}^2 - 
  \frac{\alpha}{4} \left( J^2-1 \right)^2 \, .
\end{eqnarray}
The rotational symmetry in transverse space implies that the
$(\theta\,\theta)$ and the $(\varphi\varphi)$ components  of the
Einstein equations are identical (and that
$\epsilon_\varphi=\epsilon_\theta$). Equations  (\ref{EinsteinP0}) -
(\ref{EinsteinPtheta}) are not functionally independent \cite{GRS}.
They are related by the  Bianchi identities (or equivalently by
conservation of stress-energy $\nabla_M T^{M}_{N}=0$).
Following the lines of \cite{GRS} we can define various components of
the brane tension per unit length  by 
\begin{equation} 
\label{branetensions}
  \mu_i = -\int\limits_0^{\infty} dr M(r)^4 \mathcal{L}(r)^2
\epsilon_i(r) \, .
\end{equation}

The Ricci scalar and the curvature invariants $R$, $R_{A B}\,R^{A
 B}$, $R_{A B C D}\,R^{A B C D}$,  $C_{A B C D}\,C^{A B C D}$ with 
 $C_{A B C D}$ being the Weyl tensor are given by
\begin{eqnarray}
    \frac{R}{m_W^2}&=&
    \frac{2}{{\mathcal{L}}^2} - 
    \frac{2\,{\mathcal{L}'}^2}{{\mathcal{L}}^2} - 
    \frac{16\,\mathcal{L}'\,M'}{\mathcal{L}\,M} - 
    \frac{12\,{M'}^2}{{M}^2} - 
    \frac{4\,\mathcal{L}''}{\mathcal{L}} - 
    \frac{8\,M''}{M} \, ,
    \label{Ricci} \\
   \frac{R_{A B}\,R^{A B}}{m_W^4}&=&
    \frac{2}{{\mathcal{L}}^4} - 
    \frac{4\,{\mathcal{L}'}^2}{{\mathcal{L}}^4} + 
    \frac{2\,{\mathcal{L}'}^4}{{\mathcal{L}}^4} - 
    \frac{16\,\mathcal{L}'\,M'}{{\mathcal{L}}^3\,M} + 
    \frac{16\,{\mathcal{L}'}^3\,M'}{{\mathcal{L}}^3\,M} + 
    \frac{48\,{\mathcal{L}'}^2\,{M'}^2}{{\mathcal{L}}^2\,{M}^2}+
     \nonumber \\
    &\phantom{=}&\frac{48\,\mathcal{L}'\,{M'}^3}{\mathcal{L}\,{M}^3}+
    \frac{36\,{M'}^4}{{M}^4} -
    \frac{4\,\mathcal{L}''}{{\mathcal{L}}^3} + 
    \frac{4\,{\mathcal{L}'}^2\,\mathcal{L}''}{{\mathcal{L}}^3} + 
    \frac{16\,\mathcal{L}'\,M'\,\mathcal{L}''}{{\mathcal{L}}^2\,M} + 
    \frac{6\,{\mathcal{L}''}^2}{{\mathcal{L}}^2} +\nonumber\\
    &\phantom{=}&\frac{16\,\mathcal{L}'\,M'\,M''}{\mathcal{L}\,{M}^2}
   +\frac{24\,{M'}^2\,M''}{{M}^3} + 
    \frac{16\,\mathcal{L}''\,M''}{\mathcal{L}\,M} + 
    \frac{20\,{M''}^2}{{M}^2} \, ,
    \label{RabRab} \\
   \frac{R_{A B C D}\,R^{A B C D}}{m_W^4}&=&
    \frac{4}{{\mathcal{L}}^4} - 
    \frac{8\,{\mathcal{L}'}^2}{{\mathcal{L}}^4} + 
    \frac{4\,{\mathcal{L}'}^4}{{\mathcal{L}}^4} + 
    \frac{32\,{\mathcal{L}'}^2\,{M'}^2}{{\mathcal{L}}^2\,{M}^2} + 
    \frac{24\,{M'}^4}{{M}^4} + 
    \frac{8\,{\mathcal{L}''}^2}{{\mathcal{L}}^2} + 
    \frac{16\,{M''}^2}{{M}^2} \, ,
    \label{RabcdRabcd} \\
   \frac{C_{A B C D}\,C^{A B C
    D}}{m_W^4}&=&\frac{8}{3\,{\mathcal{L}}^4} - 
    \frac{16\,{\mathcal{L}'}^2}{3\,{\mathcal{L}}^4} + 
    \frac{8\,{\mathcal{L}'}^4}{3\,{\mathcal{L}}^4} + 
    \frac{128\,\mathcal{L}'\,M'}{15\,{\mathcal{L}}^3\,M} - 
    \frac{128\,{\mathcal{L}'}^3\,M'}{15\,{\mathcal{L}}^3\,M} - 
    \frac{16\,{M'}^2}{5\,{\mathcal{L}}^2\,{M}^2} + \nonumber \\ 
    &\phantom{=}&\frac{208\,{\mathcal{L}'}^2\,{M'}^2}
            {15\,{\mathcal{L}}^2\,{M}^2}- 
    \frac{64\,\mathcal{L}'\,{M'}^3}{5\,\mathcal{L}\,{M}^3} + 
    \frac{24\,{M'}^4}{5\,{M}^4} + 
    \frac{32\,\mathcal{L}''}{15\,{\mathcal{L}}^3} - 
    \frac{32\,{\mathcal{L}'}^2\,\mathcal{L}''}{15\,{\mathcal{L}}^3} -
    \nonumber \\ 
    &\phantom{=}&\frac{64\,\mathcal{L}'\,M'\,
         \mathcal{L}''}{15\,{\mathcal{L}}^2\,M}
   +\frac{32\,{M'}^2\,\mathcal{L}''}{5\,\mathcal{L}\,{M}^2} +
    \frac{64\,{\mathcal{L}''}^2}{15\,{\mathcal{L}}^2}-
    \frac{32\,M''}{15\,{\mathcal{L}}^2\,M} + 
    \frac{32\,{\mathcal{L}'}^2\,M''}{15\,{\mathcal{L}}^2\,M} + 
    \nonumber \\
    &\phantom{=}&\frac{64\,\mathcal{L}'\,M'\,M''}{15\,
          \mathcal{L}\,{M}^2} - 
    \frac{32\,{M'}^2\,M''}{5\,{M}^3} - 
    \frac{128\,\mathcal{L}''\,M''}{15\,\mathcal{L}\,M} + 
    \frac{64\,{M''}^2}{15\,{M}^2} \, .
    \label{CabcdCabcd}
\end{eqnarray}
They must be finite continuous functions for regular geometries we are
interested in.

\section{Boundary conditions and asymptotics of the solutions} 
\label{BC}
The boundary conditions should lead to a regular solution at the
origin. Thus we have to impose
\begin{equation} 
\label{BCMorigin}
  M\vert_{r=0}=1 \, , \quad   M'\vert_{r=0}=0 \, , \quad  
  \mathcal{L}\vert_{r=0}=0 \, , \quad \mathcal{L}'\vert_{r=0}=1 \, ,
\end{equation}
for the components of the metric, where the value $+1$ for
$M\vert_{r=0}$ is a convenient choice that can be obtained by
rescaling of the brane coordinates.

What concerns the gauge and the scalar fields, the boundary
conditions for them are the same as for a monopole solution in the
flat space-time, \cite{tHooft}:
\begin{equation} 
  J(0)=0 \, , \quad  \lim_{r \to \infty} J(r) = 1 \, ,
\end{equation} 
\begin{equation} 
\label{Kboundary}
 K(0)=1 \, , \quad  \lim_{r \to \infty} K(r) = 0.
\end{equation}
Finally, a requirement of gravity localization reads
\begin{equation} 
\label{PlanckMass}
 \frac{4 \pi M_7^5}{m_W^3} \int\limits_{0}^{\infty} M(r)^2
\mathcal{L}(r)^2 dr = M_P^2 < \infty~,
\end{equation}
what puts a constraint on the behavior of the metric at infinity.

The magnetic charge of the field configuration can either be
determined by comparing the stress-energy tensor with  the general
expression given in \cite{GRS} or by a direct calculation of the
magnetic field strength tensor \cite{tHooft}:
\begin{equation} 
\label{FST} \mathcal{G}_{M N}=\frac{\vec{\Phi} \cdot \vec{G}_{M
N}}{\vert \vec{\Phi}\vert}- \frac{1}{e \vert \vec{\Phi}\vert^3} \,
\vec{\Phi} \cdot \left( \mathcal{D}_M \vec{\Phi} \times
\mathcal{D}_N  \vec{\Phi} \right).
\end{equation}
The only nonzero component of $ \mathcal{G}_{M N}$ is $
\mathcal{G}_{\theta \varphi}=-\frac{\sin\theta}{\vert e \vert}$.
Either way gives $Q=\frac{1}{e}$.

\subsection{Behavior at the origin} \label{BSO}
Once boundary conditions at the center of the defect are imposed for
the fields and the metric, the system of equations
(\ref{EinsteinP0})-(\ref{EqMovW}) can be solved in the vicinity of
the origin by developing the fields and the  metric into a power
series in the (reduced) transverse radial variable $r$. For the given
system this can be done up to any desired order. We give the power
series up to third order in $r$:
\begin{eqnarray}
  M(r)  \hspace{-2mm} &=& \hspace{-2mm} 1 - \frac{1}{60} r^2\,\beta
 \,\left( \alpha  + 4\,\gamma  - 6\,{K''(0)}^2 \right) +
 \mathcal{O}(r^4) \, ,  
 \label{asy_origin_M}\\ 
 \mathcal{L}(r)\hspace{-2mm} &=& \hspace{-2mm} r + \frac{1}{360}
 r^3\,\beta \,\left( \alpha  + 4\,\gamma - 30\,{J'(0)}^2 -
 66\,{K''(0)}^2 \right) + \mathcal{O}(r^5) \, , 
 \label{asy_origin_L}\\ J(r)\hspace{-2mm} &=&  \hspace{-2mm} r J'(0)
 + r^3\,J'(0) \frac{-9\alpha  + \alpha \beta  + 4\beta \gamma  +
 6\beta {J'(0)}^2 + \hspace{-1mm} 18K''(0) + 6\beta {K''(0)}^2}{90}
 \hspace{-1mm}+ \mathcal{O}(r^5) \, ,  \label{asy_origin_J}\\ K(r)
 \hspace{-2mm} &=&\hspace{-2mm} 1 + \frac{1}{2} r^2\,K''(0) +
 \mathcal{O}(r^4) \,  
 \label{asy_origin_K}.
\end{eqnarray}

It can easily be shown that the power series of  $M(r)$ and $K(r)$
only involve even powers of $r$ whereas those of  $\mathcal{L}(r)$
and $J(r)$ involve only odd ones. The expressions for $\mathcal{L}(r)
$  and $J(r)$ are therefore valid up  to $5^{\mathrm th}$ order. One
observes that the solutions satisfying the boundary  conditions at
the origin can be parametrized by five parameters
$\left(\alpha,\beta,\gamma, J'(0), K''(0) \right)$. For arbitrary
combinations of these parameters the corresponding metric solution
will not satisfy the boundary  conditions at infinity. Therefore the
task is to find those parameter combinations for which 
(\ref{PlanckMass}) is finite. For completeness, we give the zero-th
order of the power series solutions for the stress-energy tensor
components and the curvature invariants at the origin:
\begin{eqnarray} 
\label{epsorigin}
  \epsilon_0 \vert_{r=0}&=&-\frac{1}{4} \left( \alpha + 6 \, J'(0)^2
+ 6 \, K''(0)^2 \right) \, , \\ 
\epsilon_{\rho}\vert_{r=0}=\epsilon_{\theta}\vert_{r=0}&=&-\frac{1}{4}
\left( \alpha + 2 \, J'(0)^2 - 2 \, K''(0)^2 \right). 
\end{eqnarray}
\begin{eqnarray} 
\label{curvatorigin} 
\frac{R}{m_W^2}&=&
\frac{\beta}{10} \, \left( 7\,\alpha  + 28\,\gamma  + 30\,{J'(0)}^2 +
18\,{K''(0)}^2 \right)\, ,\\ 
\frac{R_{A B}\,R^{A
B}}{m_W^4}&=&\frac{{\beta}^2}{100} \,\left( 7\,{\alpha }^2 +
56\,\alpha \,\gamma  + 112\,{\gamma }^2 + 60\,\alpha \,{J'(0)}^2 +
240\,\gamma \,{J'(0)}^2 + 300\,{J'(0)}^4 \right. \nonumber \\
  &&
\quad \left. + 12\,\left( 3\,\alpha  + 12\,\gamma  +  70\,{J'(0)}^2
\right) \,{K''(0)}^2 + 732\,{K''(0)}^4 \right) \, ,\\ 
\frac{R_{A B C
D}\,R^{A B C D}}{m_W^4}&=&\frac{{\beta }^2}{300} \, \left(
17\,{\alpha }^2 + 136\, \alpha \, \gamma +  272 \, {\gamma}^2 - 60\,
\alpha \, {J'(0)}^2 - 240\,\gamma \, {J'(0)}^2 + 900 \, {J'(0)}^4
\right. \nonumber \\ && \quad \left. - 36\,\left( 9\,\alpha  +
36\,\gamma  -  110\,{J'(0)}^2 \right) \,{K''(0)}^2 + 4932\,{K''(0)}^4
\right)\, ,\\ 
\frac{C_{A B C D}\,C^{A B C D}}{m_W^4}&=& \frac{{\beta
}^2}{30}\,{\left( \alpha  + 4\,\gamma  - 6\,\left( {J'(0)}^2 +
3\,{K''(0)}^2 \right)  \right) }^2 \, .
\end{eqnarray}

\subsection{Behavior at infinity}\label{BSI}
The asymptotics of the metric functions $M(r)$ and $\mathcal{L}(r)$
far away from the monopole are \cite{GRS}:
\begin{equation} 
\label{asympto}
  M(r)=M_0 \, e^{-\frac{c}{2}r} \quad \mbox{and} \quad
\mathcal{L}(r)=\mathcal{L}_0=\mbox{const}\, ,
\end{equation}
where only positive values of $c$ lead to gravity localization. This
induces the following asymptotics for  the stress-energy components
and the various curvature invariants:
\begin{eqnarray} 
\label{epsinfinity}
  \lim_{r \to \infty} \epsilon_0(r)=\lim_{r \to \infty}
  \epsilon_{\rho}(r)&=&-\frac{1}{2 {\mathcal{L}_0}^4} \, , \\ \lim_{r
 \to \infty} \epsilon_{\theta}(r)&=&\frac{1}{2 {\mathcal{L}_0}^4} \,.
\end{eqnarray}
\begin{eqnarray}
  \lim_{r \to \infty} \frac{R(r)}{m_W^2} &=& -5 c^2 +
  \frac{2}{\mathcal{L}_0^2} \, , 
  \label{curvinvinfinity1}\\ \lim_{r
  \to \infty} \frac{R_{A B}(r) R^{A B}(r)}{m_W^4} &=& 5 c^4 +
  \frac{2}{\mathcal{L}_0^4} \, , 
  \label{curvinvinfinity2}\\ \lim_{r
  \to \infty} \frac{R_{A B C D}(r) R^{A B C D}(r)}{m_W^4} &=&
  \frac{5}{2} c^4 + \frac{4}{\mathcal{L}_0^4} \, ,
  \label{curvinvinfinity3}\\ \lim_{r \to \infty} \frac{C_{A B C D}(r)
  C^{A B C D}(r)}{m_W^4} &=& \frac{c^4}{6} +\frac{8}{3 \,
  {\mathcal{L}_0^4}} -\frac{4 c^2}{\mathcal{L}_0^2} \, .
  \label{curvinvinfinity4}
\end{eqnarray}
The parameters $c$ and $\mathcal{L}_0$ are determined  by Einsteins
equations for large $r$ and are given by \cite{GRS}:
\begin{eqnarray} 
  c^2&=&\frac{5}{32 \beta} \left( 1-\frac{16}{5} \gamma \beta^2  \pm
  \sqrt{1-\frac{32}{25} \gamma \beta^2} \right)\, , 
  \label{Einsteinasymptocsqr}\\
  \frac{1}{\mathcal{L}_0^2}&=&\frac{5}{8\beta}\left( 1 \pm
  \sqrt{1-\frac{32}{25} \gamma \beta^2} \right) 
   \label{Einsteinasympto1overLsqr} \, .
\end{eqnarray}
Only the positive signs of the roots lead to solutions with both
$c^2>0$ and $\frac{1}{\mathcal{L}_0^2}>0$ \cite{GRS}. 

In order to obtain some information about the asymptotic behavior of
the fields $J(r)$ and $K(r)$ at infinity we insert relations
(\ref{asympto}) into (\ref{EqMovphi}) and (\ref{EqMovW}). Furthermore
we use $K(r)=\delta K(r)$ and  $J(r)=1-\delta J(r)$ with $\delta K
\ll 1$ and $\delta J \ll 1$ for $1\ll r$ to linearize these
equations:
\begin{eqnarray}
  \delta J'' - 2 \, c \, \delta J' -2 \alpha \delta J &=& 0 \, ,
  \label{lineqJ}\\ \delta K'' - 2 \, c \, \delta K'
  +\left(\frac{1}{\mathcal{L}_0^2}-1\right) \delta K &=& 0\,
  \label{lineqK} .
\end{eqnarray}
By using the ansatz $\delta K = A e^{-k r}$ and $\delta J = B e^{-j
r}$ we find 
\begin{eqnarray}
  k_{1,2}=-c\pm\sqrt{c^2-\left( \frac{1}{\mathcal{L}_0^2}-1\right)} \, ,
  \label{kasympt} \qquad  j_{1,2}=-c\pm\sqrt{c^2+2\alpha} \, .
\end{eqnarray} 
To satisfy the boundary conditions we obviously have to impose $k>0$
and $j>0$. We distinguish two cases:
\begin{enumerate}
\item \fbox{$c > 0$} Gravity localizing solutions. In this case there
is a unique  $k$ for $\frac{1}{\mathcal{L}_0^2}<1$ with  the positive
sign in (\ref{kasympt}).
\item \fbox{$c < 0$} Solutions that do not localize gravity.
\begin{enumerate}
  \item \fbox{$\frac{1}{\mathcal{L}_0^2} > 1 $}  
 \raisebox{-5mm}{
  \parbox{13cm}{
  \begin{enumerate}
    \item \fbox{$c^2 \geq \frac{1}{\mathcal{L}_0^2}-1$} \, Both
    solutions of (\ref{kasympt}) are positive. \item \fbox{$c^2 <
    \frac{1}{\mathcal{L}_0^2}-1$} \, In this case there are no real
    solutions.
  \end{enumerate}
  }
  }
  \item \fbox{$\frac{1}{\mathcal{L}_0^2} < 1 $} \, There is a unique
solution with the positive sign in (\ref{kasympt}) .
  \item \fbox{$\frac{1}{\mathcal{L}_0^2} = 1 $} \, The important
$k$-value here is $k=-2 c$.
\end{enumerate}
\end{enumerate}
It can be easily shown  that for large enough $r$ the linear
approximation (\ref{lineqK})  to the equation of motion of the gauge
field is always valid. Linearizing the equation for the scalar field
however breaks down for $2k<j$ due to the presence of the term
$\propto K^2 J$ in (\ref{EqMovphi}). In that case $J(r)$ approaches
$1$ as $e^{-2 k r}$. For solutions with $c>0$ (the case of 
predominant interest) a detailed discussion of the validity of $2k>j$
gives:
\begin{enumerate} 
\item \fbox{$\alpha=0$} \, Prasad-Sommerfield limit
\cite{PraSom}. One has $2k > j$. The asymptotics of the scalar field
is  governed by $e^{- j r}$.
\item \fbox{$0<\alpha<2$} \, The validity of $2k > j$ depends on
 different inequalities between $\alpha$, $\frac{1}{\mathcal{L}_0^2}$
 and \nolinebreak $c^2$.
  \begin{enumerate}
    \item \fbox{$\alpha+\frac{1}{\mathcal{L}_0^2} \leq 1$}  $
\longrightarrow 2k > j $ \, .
    \item  \fbox{$\alpha+\frac{1}{\mathcal{L}_0^2} > 1$} and 
     \fbox{$\frac{\alpha}{2}+\frac{1}{\mathcal{L}_0^2} < 1$}\, . In
     this case $ 2k \geq j$ is equivalent to  
     $c^2 \leq \frac{-\left(1 - \frac{\alpha}{2} -
     \frac{1}{\mathcal{L}_0^2}\right)^2}
     {1-\alpha-\frac{1}{\mathcal{L}_0^2}}$,
     where equality in one of the equations implies equality in the
     other.  
    \item \fbox {$\frac{\alpha}{2}+\frac{1}{\mathcal{L}_0^2} \geq
    1$}  $ \longrightarrow 2k < j $ and the scalar field  asymptotics
    is governed by $e^{- 2 k r}$.
  \end{enumerate}
\item \fbox{$\alpha >2$} $ \longrightarrow 2k < j $. Also in this
  case the gauge field $K(r)$ determines  the asymptotics of the
  scalar field $J(r)$.
\end{enumerate}

\subsection{Fine-tuning relations} \label{FTR}
It is possible to derive analytic relations between the different
components of the brane tensions valid for  gravity localizing
solutions. Integrating linear combinations of Einsteins equations 
(\ref{EinsteinP0})-(\ref{EinsteinPtheta}) between $0$ and $\infty$ 
after multiplication with $M^4(r) \mathcal{L}^2(r)$ gives:
\begin{eqnarray}
  \mu_0-\mu_{\theta}&=&\frac{1}{\beta} \int\limits_{0}^{\infty} M^4 d
  r =  \int\limits_{0}^{\infty} \left(1 -K^2
  \right)\frac{M^4}{\mathcal{L}^2} d r \, , \label{FineTunRel1}\\
  \mu_0-\mu_{\rho}- 2 \mu_{\theta}&=&2 \gamma
  \int\limits_{0}^{\infty} M^4 \mathcal{L}^2 d r \, ,
  \label{FineTunRel2}\\ 
  \mu_0+\mu_{\rho}+ 2 \mu_{\theta}&=&  \alpha
  \int\limits_{0}^{\infty} \left( 1 - J^2 \right)  M^4 \mathcal{L}^2
  d r \, 
  \label{FineTunRel3}.
\end{eqnarray} 
To obtain the above relations integration by parts was used where the
boundary terms dropped due to the boundary conditions given above. A
detailed derivation of these relations is given in the appendix A.

\newpage
\section{Numerical solutions} \label{N}

Details of numerical integrations are given in the Appendix
\ref{ApB}; in this section we will discuss the results only.

\subsection{Examples of numerical solutions} \label{samplesol}

\begin{figure}[hb]
\centerline{ \epsfxsize = 15cm \epsfbox {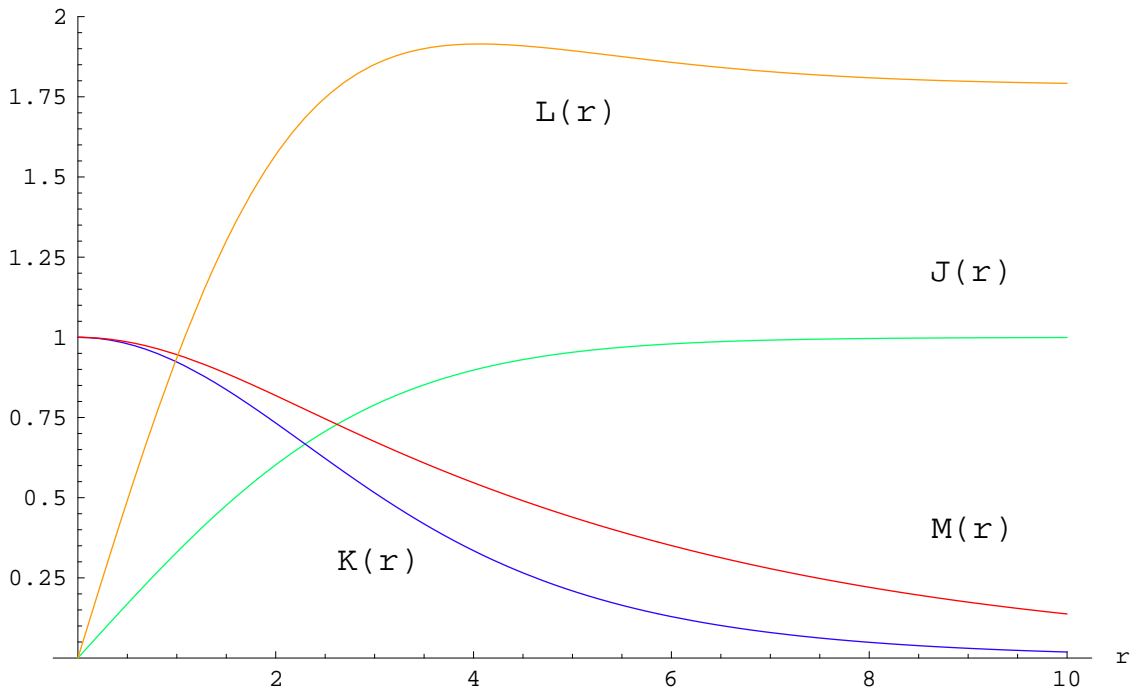}}  
\caption[a]{\it Gravity-localizing solution with negative bulk
cosmological constant corresponding to the parameter values  
$(\alpha=1.00000000, \, \beta=5.50000000, \, \gamma=-0.05434431) $}
\label{sol155}
\end{figure}

\begin{figure}
\centerline{ \epsfxsize = 15cm \epsfbox {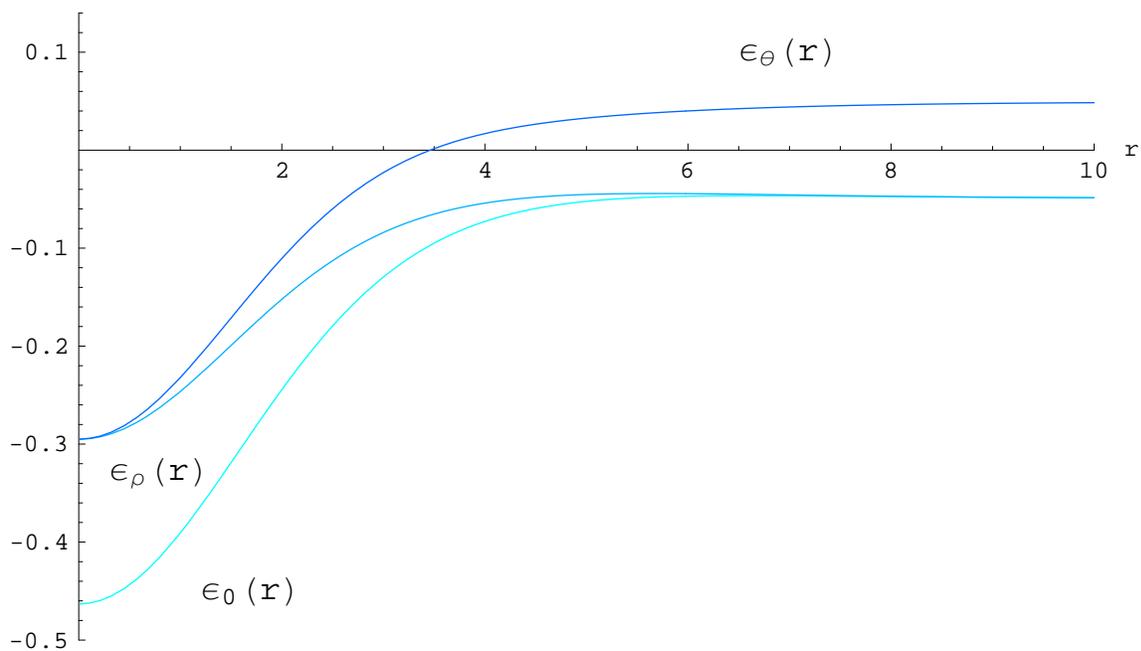}}
\caption[a]{\it Stress-energy components for the solution given in
Fig.~\ref{sol155}. \\ $(\alpha=1.00000000, \, \beta=5.50000000, \,
\gamma=-0.05434431)$} \label{sol155_e}
\end{figure}

\begin{figure}
\centerline{ \epsfxsize = 15cm \epsfbox {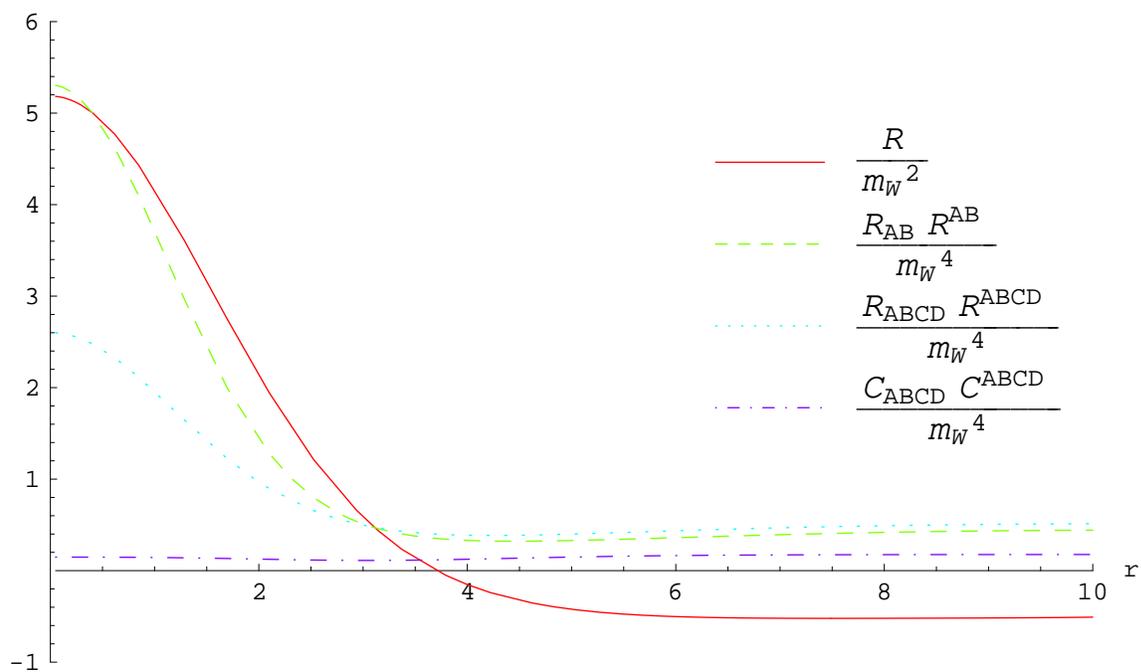}}
\caption[a]{\it Curvature invariants for the solution given in
Fig.~\ref{sol155}. \\ $(\alpha=1.00000000, \, \beta=5.50000000, \,
\gamma=-0.05434431) $} \label{sol155_c}
\end{figure}

\begin{figure}
\centerline{ \epsfxsize = 15cm \epsfbox {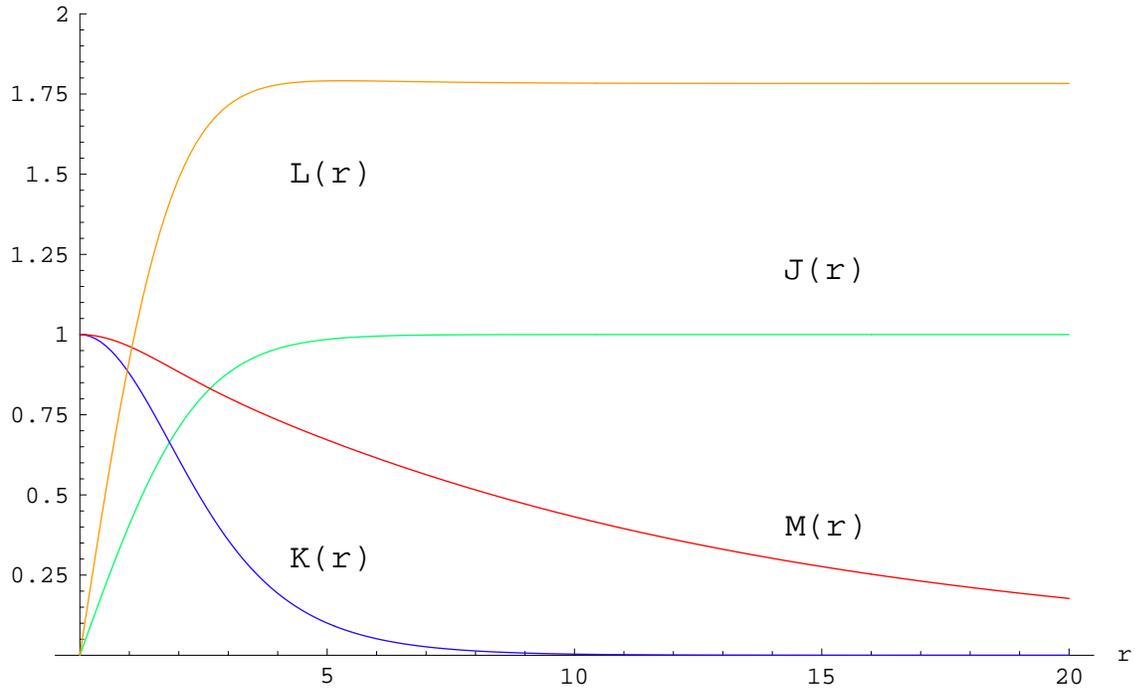}}
\caption[a]{\it Gravity-localizing solution with positive bulk
cosmological constant corresponding to the parameter values
$(\alpha=1.00000000, \, \beta=3.50000000, \, \gamma=0.02678351) $}
\label{sol135}
\end{figure}

\begin{figure}
\centerline{ \epsfxsize = 15cm \epsfbox {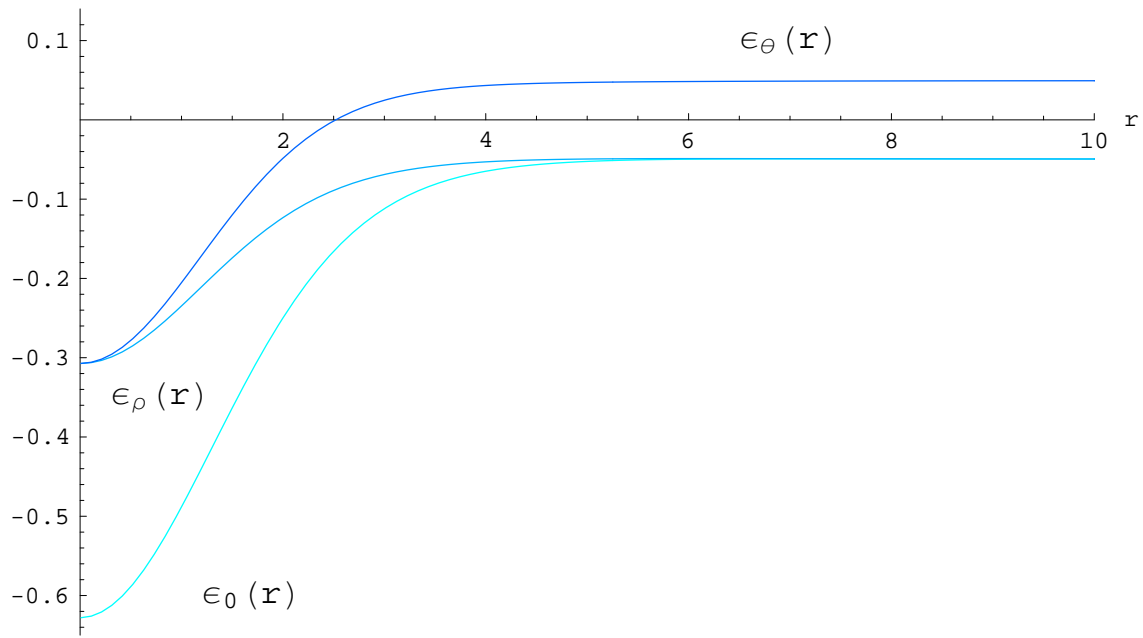}}
\caption[a]{\it Stress-energy components for the solution given in
Fig.~\ref{sol135}. \\ $(\alpha=1.00000000, \, \beta=3.50000000, \,
\gamma=0.02678351)$} \label{sol135_e}
\end{figure}

\begin{figure}
\centerline{ \epsfxsize = 15cm \epsfbox {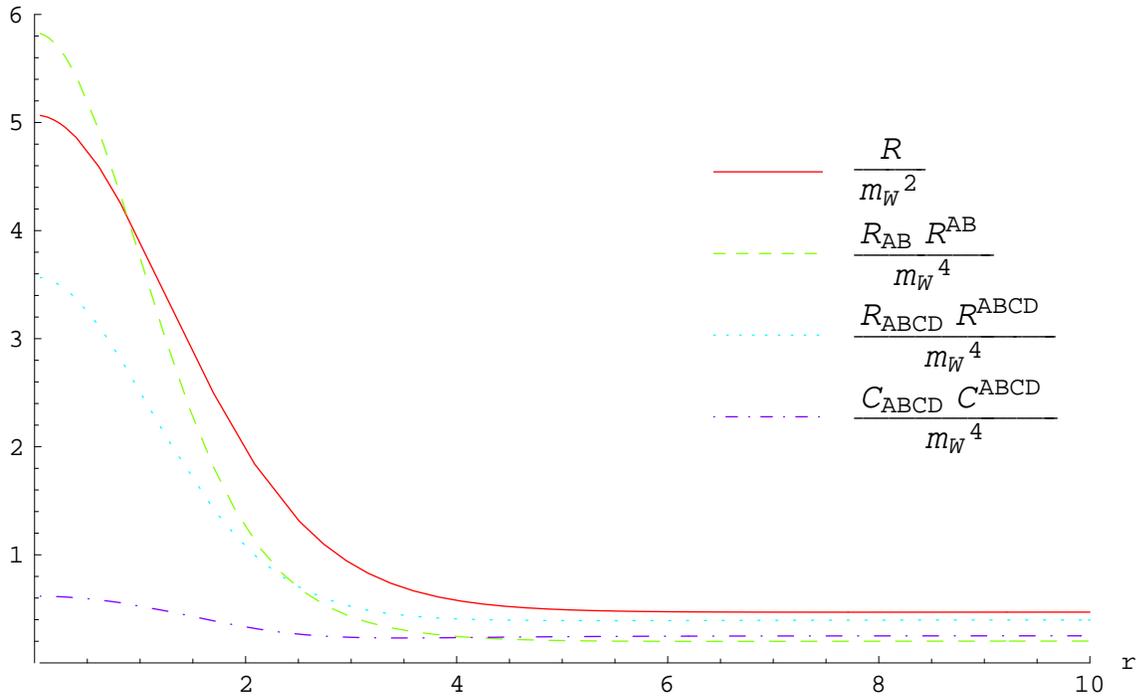}}
\caption[a]{\it Curvature invariants for the solution given in
Fig.~\ref{sol135}. \\ $(\alpha=1.00000000, \, \beta=3.50000000, \,
\gamma=0.02678351) $}\label{sol135_c}
\end{figure}

\begin{figure}
\centerline{ \epsfxsize = 15cm \epsfbox {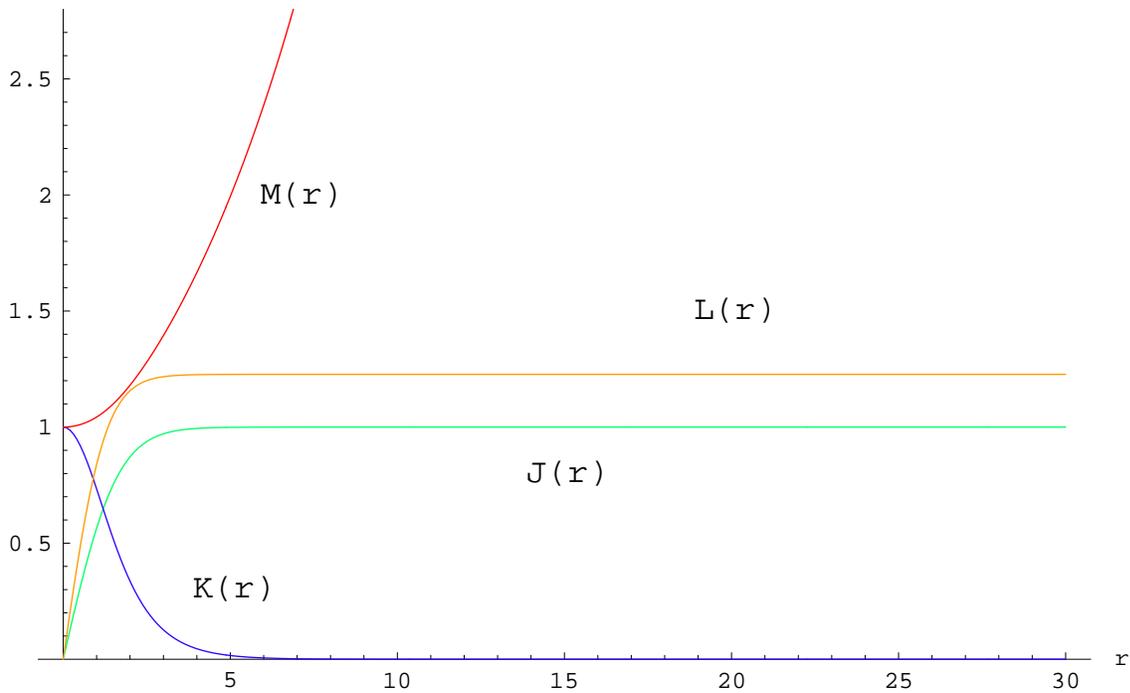}}
\caption[a]{\it Example of a solution that does not localize gravity
$(c<0)$ corresponding to the parameter values $(\alpha=1.00000000, \,
\beta=1.80000000, \, \gamma=0.04053600) $} \label{sol118}
\end{figure}

\begin{figure}
\centerline{ \epsfxsize = 15cm \epsfbox {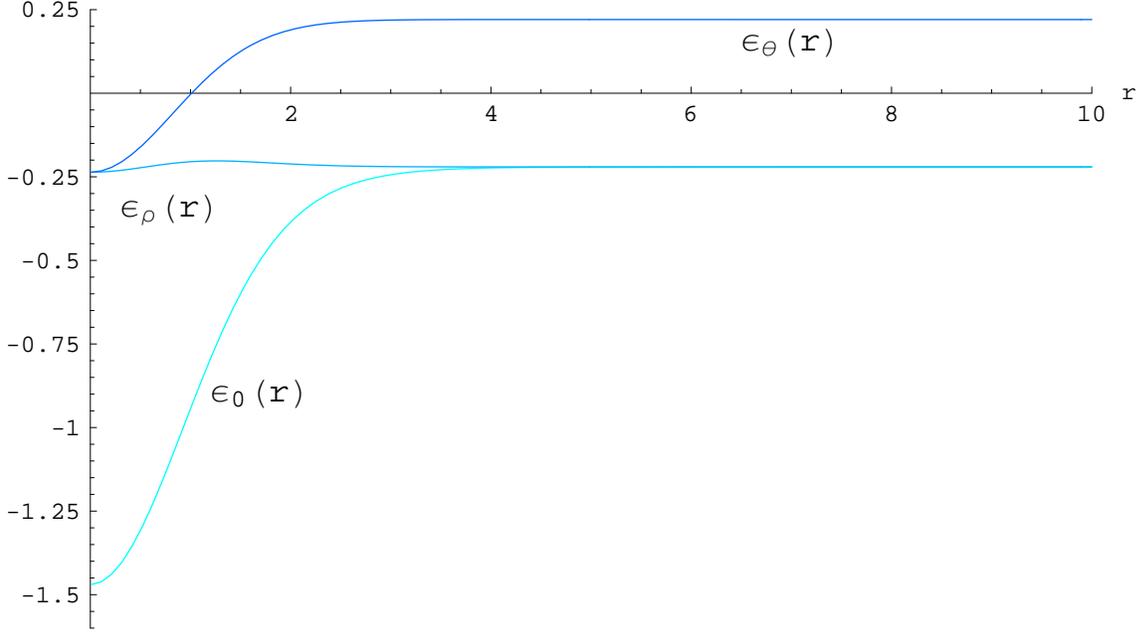}}
\caption[a]{\it Stress-energy components for the solution given in
Fig.~\ref{sol118}. \\ $(\alpha=1.00000000, \, \beta=1.80000000, \,
\gamma=0.04053600)$} \label{sol118_e}
\end{figure}

Fig.~\ref{sol155} and Fig.~\ref{sol135} show two different numerical
solutions corresponding to positive  and negative bulk cosmological
constant, respectively. Both solutions localize gravity $c>0$. The
corresponding parameter values are  $(\alpha=1.0000000 , \,
\beta=5.50000000 , \, \gamma=-0.05434431) $ and  $(\alpha=1.0000000 ,
\, \beta=3.50000000 , \, \gamma=0.02678351) $. Figs.~\ref{sol155_e}
and~\ref{sol135_e} and Figs.~\ref{sol155_c} and~\ref{sol135_c} show
the corresponding components of the stress-energy tensor and the
curvature invariants. For high values of $\beta$ the metric function
$\mathcal{L}(r)$ develops a maximum before attaining its boundary
value. Gravity dominates and the volume of the transverse space stays
finite.  For lower values of $\beta$ the transverse space has
infinite volume and gravity can not be localized.  See
Figs.~\ref{sol118} and ~\ref{sol118_e} for a solution that does not
localize gravity $(c<0)$ and corresponds to  $(\alpha=1.0000000 , \,
\beta=1.80000000 , \, \gamma=0.04053600) $.
\clearpage
\clearpage
\clearpage
\clearpage

\subsection{The fine-tuning surface} \label{NFTS}
\begin{figure}
\centerline{ \epsfxsize = 10cm \epsfbox {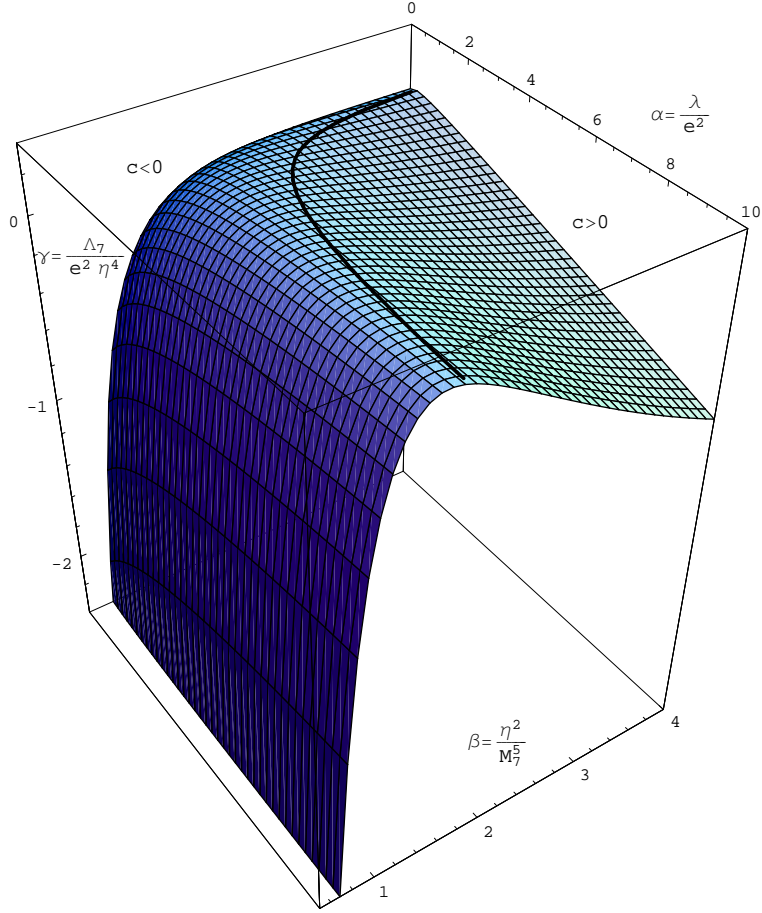}}
\caption[a]{\it Fine-tuning surface for solutions with the metric
asymptotics (\ref{asympto}). The bold line separates solutions that
localize gravity $(c>0)$ from those that do not $(c<0)$. Numerically
obtained values of  $\gamma=\frac{\Lambda_7}{e^2 \eta^4}$ are plotted
as a function of $\alpha = \frac{\lambda}{e^2}$ and  $\beta =
\frac{\eta^2}{M_7^5}$.}
\label{FTS}
\end{figure}

Fig.~\ref{FTS} shows the fine-tuning surface in  parameter-space.
A point on this surface $(\alpha_0,\,\beta_0,\,\gamma_0)$ corresponds
to a particular solution with the metric asymptotics (\ref{asympto}) 
for both values of the sign of $c$.
The bold line separates gravity localizing solutions $c>0$ from
solutions that do not localize gravity $c<0$. The parameter space has
been thoroughly exploited within the rectangles  $\Delta\alpha \times
\Delta\beta = [0,\,10]\,  \times \, [0.5,\, 4.0]$ and  $\Delta\alpha
\times \Delta\beta = [0,\,1]\,  \times \, [3.5,\, 10]$  in the
$(\alpha,\beta)$-plane. The series of solutions presented in section
\ref{samplesol} can be used to illustrate their dependence on the 
parameter $\beta$ (strength of gravity) for a for a fixed value of 
$\alpha=1$.

It can be seen from Fig.~\ref{FTS} that for every fixed $\alpha$
there is a particular value of  $\beta$ such that $c$ equals zero,
which is the case for all points on the solid line shown in
Fig.~\ref{FTS}. By  looking at eq.~(\ref{Einsteinasymptocsqr}) we
immediately see that $c=0$ is equivalent to $\beta^2
\gamma=\frac{\Lambda_7}{e^2 \, M_7^{10}}=\frac{1}{2}$.  We will
discuss the limit $c \ll 1$ in more detail in section \ref{FML}. It will
turn out to be the most physical case where the monopole can be 
considered to be point-like since the fields attain their vacuum 
values much earlier than the metric goes to zero outside the core.  
A solution corresponding to that 
case is given in Fig.~\ref{solFML}, where gravity 
(parametrized  by $\beta$)
is just strong enough to provide a finite volume for transverse space.
If for fixed value of $\alpha$ we increase $\beta$ (starting from
$c=0$), $c$ becomes more
and more positive, the Planck mass 
becomes smaller and the monopole size increases, see 
Figs.~\ref{sol155} and \ref{sol135}. 
If on the other hand $\beta$ is decreased (from $c=0$ on), 
$c$ becomes more
and more negative and the metric $M(r)$ blows up exponentially, as 
in Fig.~\ref{sol118}. Gravity is no longer strong enough to provide
for a finite Planck mass.

For $\alpha$=0 (the Prasad-Sommerfield limit, to be discussed 
in section \ref{PSL}) it can be read off from Fig.~\ref{FTS} that 
there are no solutions that localize gravity. All $\alpha=0$ solutions 
lie in the $c<0$ part of the surface.

\section{Prasad-Sommerfield limit $(\alpha=0)$} \label{PSL}

\begin{figure}[htbp]
\centerline{\epsfxsize = 12cm \epsfbox {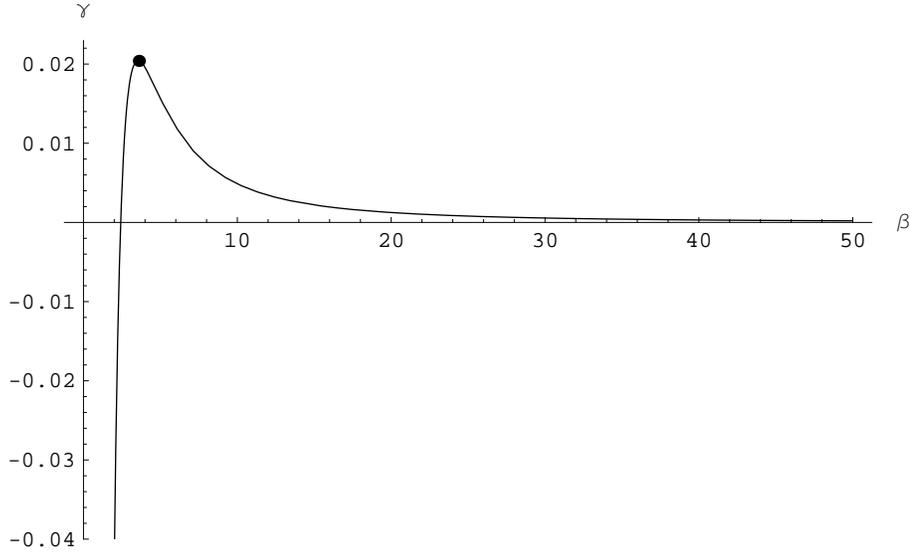}}
\caption[a]{\it Section of the fine-tuning surface Fig.~\ref{FTS}
corresponding to the Prasad-Sommerfield limit $\alpha=0$.  None of
the shown combinations of $\beta$ and $\gamma$-values correspond to
solutions that lead to warped compactification. The point indicates
the sample solution given below.}
\label{PSlimit}
\end{figure}

\begin{figure}[htbp]
\centerline{ \epsfxsize = 12cm \epsfbox {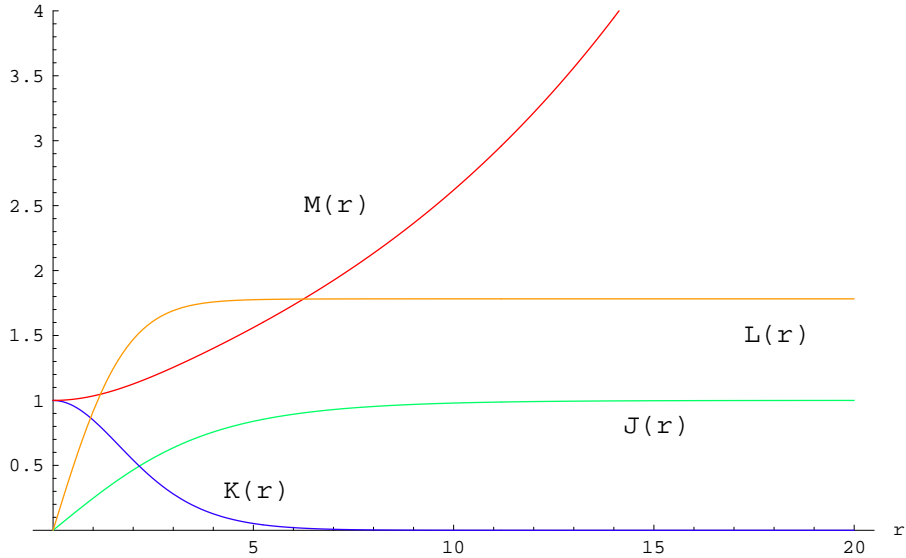}}
\caption[a]{\it Sample solution for the Prasad-Sommerfield limit
corresponding to the parameter values   $(\alpha=0.00000000, \,
\beta=3.60000000, \, \gamma=0.02040333) $}
\label{sol036}
\end{figure}

\begin{figure}[htbp]
\centerline{ \epsfxsize = 14cm \epsfbox {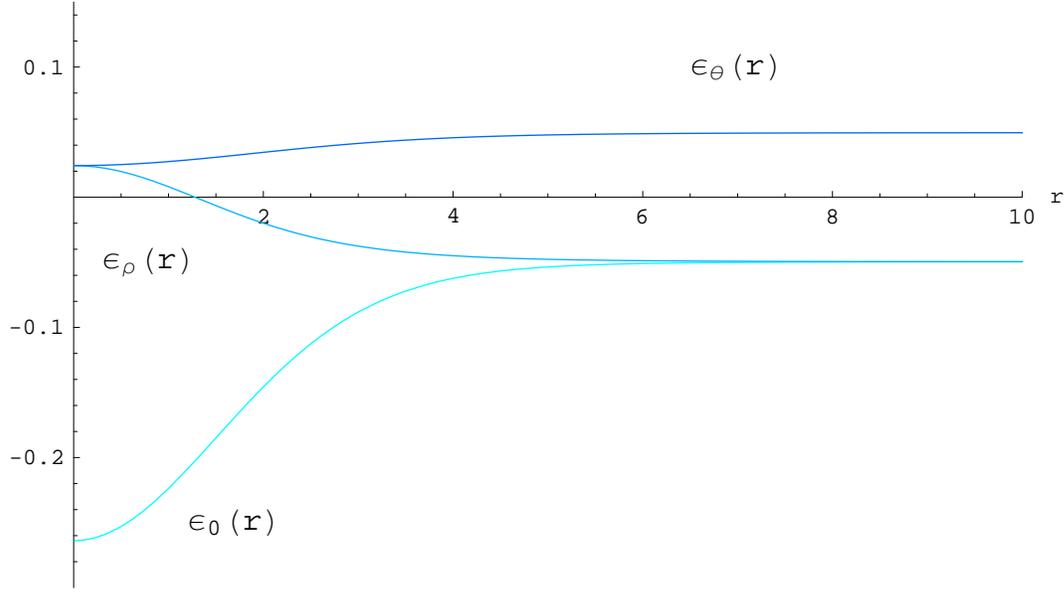}}
\caption[a]{\it Stress-energy components for the solution given in
Fig.~\ref{sol036}.\\ $(\alpha=0.00000000, \, \beta=3.60000000, \,
\gamma=0.02040333)$}
\label{sol036_e}
\end{figure}

The Prasad-Sommerfield limit $(\alpha=0)$ was exploited numerically
for $\beta$-values ranging from $0.4$ to  about $70$. The
corresponding intersection of the fine-tuning surface Fig.~\ref{FTS}
and the plane $\alpha=0$  is given in Fig.~\ref{PSlimit}. There exist
no gravity localizing solutions as the separating line in 
Fig.~\ref{FTS} indicates. The point in Fig.~\ref{PSlimit} corresponds
to the solution shown in Figs.~\ref{sol036} and \ref{sol036_e}. One
sees that $\gamma$ tends to zero for $\beta$ going to infinity and
that $\gamma$ tends to $-\infty$ for $\beta$ going to zero. 

\section{A point-like monopole limit \, - \, physical requirements on
solutions}\label{FML}

\begin{figure}[htbp]
\centerline{ \epsfxsize = 14cm \epsfbox {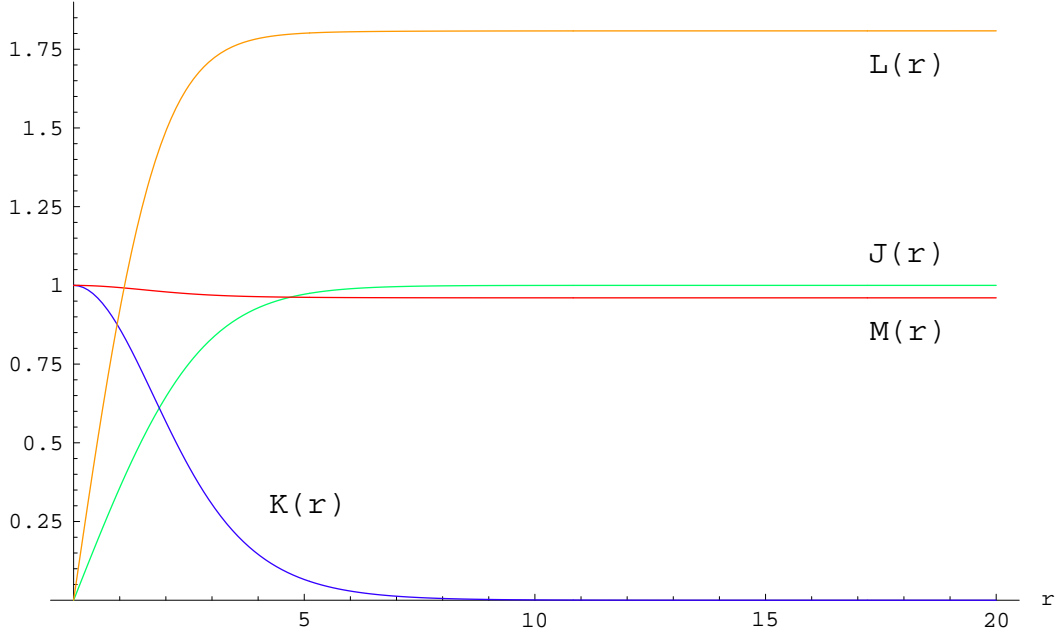}}  
\caption[a]{\it Gravity-localizing solution in the fine-monopole limit for 
$c=1.728 \cdot 10^{-5}$ and corresponding parameter values of  
$(\alpha=0.50000000, \, \beta=3.27000000, \, \gamma=0.04676000) $}
\label{solFML}
\end{figure}

The fine monopole can be characterized by $c \ll 1$. As anticipated
in section \ref{NFTS}, $c=0$  in eq.~(\ref{Einsteinasymptocsqr})
immediately leads to   $\beta^2 \gamma=\frac{\Lambda_7}{e^2 \,
M_7^{10}}=\frac{1}{2}$. Hence we deduce that the fine-monopole limit
can not be realized for a negative bulk cosmological constant. In
addition, from (\ref{Einsteinasympto1overLsqr}) it follows that
$\mathcal{L}_0 = \sqrt{\beta}$.  The $c\ll 1$ limit is qualitatively
different from its analogue in the $6D$-string case \cite{Harvey}.
The  solutions do not correspond to strictly local defects. The 
Einstein equations never decouple from the field equations. Due to the
particular metric asymptotics (\ref{asympto}), the stress-energy
tensor components tend to constants at infinity in transverse space.
In the $6D$-string case the fine-string limit was realized as a
strictly local defect having  stress energy vanishing exponentially
outside the string core. Despite this difference the discussions of
the physical requirements are very similar. In the following we show
that in the fine monopole-limit the dimension-full parameters of the
system  $\Lambda_7, \,  M_7, \, m_W, \, \lambda, \, e $ can be chosen
in such a way that all of the following physical requirements are
simultaneously satisfied:
\begin{enumerate}
  \item $M_P^2$ equals $ (1.22 \cdot 10^{19} \mbox{GeV})^2$. 
  \label{R1}
  \item The corrections to Newtons law do not contradict latest 
  measurements.\label{R2}
  \item Classical gravity is applicable in the bulk.\label{R3}
  \item Classical gravity is applicable in the monopole core $(r=0)$.
  \label{R4}
\end{enumerate}
To find solutions with the above mentioned properties it is possible
to restrict oneself to a particular value of $\alpha$, e.g.
$\alpha=\frac{1}{2}$. This choice corresponds to equal vector and
Higgs masses $m_W=m_H$. Even though extra dimensions are infinite,
the fact that $M(r)$ decreases exponentially permits the definition
of  an effective ``size'' $r_0$ of the extra dimensions:
\begin{equation} 
\label{sizeofexdim}
  M=M_0 \, e^{-\frac{c}{2}r}=M_0 \, e^{-\frac{c \,
m_W}{2}\frac{r}{m_W}} \quad \Rightarrow \quad r_0 \equiv \frac{2}{c
\, m_W} \, .
\end{equation}
In order to solve the hierarchy problem in similar lines to
\cite{Arkani-Hamed:1998rs} we parametrize the fundamental gravity  
scale as follows:
\begin{equation} 
\label{fundgravscale}
  M_7= \kappa \, 10^3 \, \mbox{GeV} \, .
\end{equation}
$\kappa=1$ then sets the fundamental scale equal to the electroweak
scale.
\begin{enumerate}
  \item The expression for the square of the Planck mass $M_P^2$ can
 be approximated in the fine-monopole  limit by using the asymptotics
 (\ref{asympto}) for the metric in the integral (\ref{PlanckMass})
 rather than the exact (numerical) solutions. This gives
  \begin{equation}
     M_P^2 \approx \frac{4 \pi M_7^5}{m_W^3}\, M_0^2 \,
\mathcal{L}_0^2 \, \frac{1}{c} \, . 
  \end{equation}
  By using one of Einstein's equations at infinity 
\begin{equation} 
\label{EinsteininfI}
  \mathcal{L}_0^2=\frac{1}{4 c^2+2 \beta \gamma} \, ,
\end{equation}
  and by developing to lowest order in $c$ one finds
  \begin{equation}
   M_P^2 \approx \frac{4 \pi M_7^5 M_0^2}{m_W^3} \frac{1}{2 \beta
\gamma} \left(\frac{1}{c}+ \mathcal{O}(c) \right) \, .  
  \end{equation}
Numerical solutions for $c \to 0$ and $\alpha=\frac{1}{2}$ converge
to the following approximate parameter values 
\begin{eqnarray} 
\label{finemononum}
  \beta &=& 3.266281 \,\,\, ,\nonumber \\
  \gamma &=& 0.0468665 \,\,\, ,\nonumber \\
  \beta^2 \gamma &=& 0.4999995 \,\,\, ,\nonumber\\
  M_0 &=& 0.959721 \,\,\, .
\end{eqnarray}
The above values were obtained by extrapolation of solutions to
$c=0$, see Figs.~\ref{fmlbeta} to \ref{fmlM0}. Therefore the 
relative errors
are of order $10^{-6}$ which is considerably higher than average, 
relative errors from the integration  which were at least $10^{-8}$.  
\begin{figure}[htbp]
\centerline{
\parbox{9.0cm}{
$\scriptstyle \beta=\frac{\eta^2}{M_7^5}$ \\
\epsfxsize = 9.0cm \epsfbox {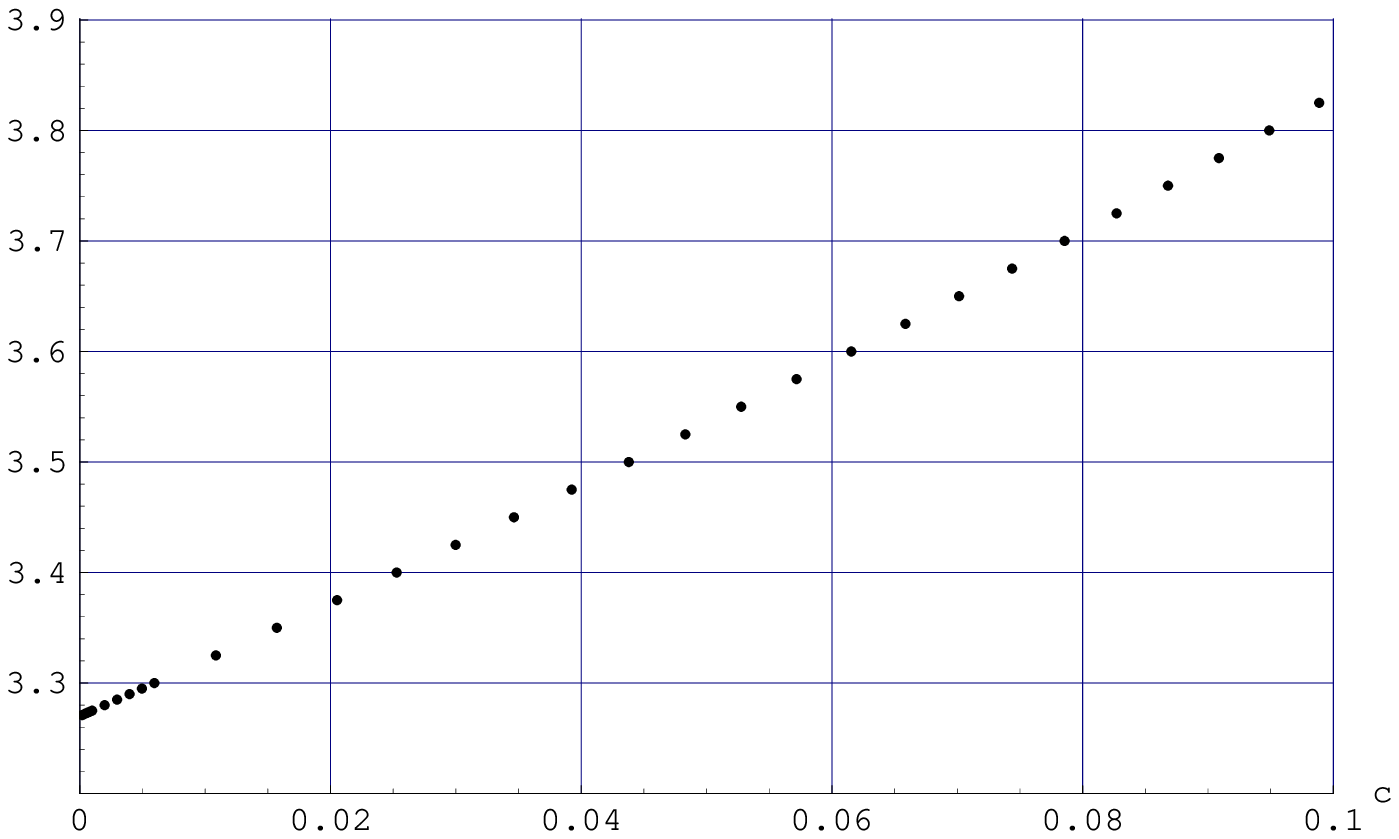}
}
}
\caption[a]{\it Behavior of $\beta$ in the fine-monopole limit $c \to
0$ for $\alpha=\frac{1}{2}$.}
\label{fmlbeta}
\end{figure}
\begin{figure}[htbp]
\centerline{
\parbox{9.0cm}{
$\scriptstyle \gamma=\frac{\Lambda_7}{e^2 \eta^4}$ \\
\epsfxsize = 9.0cm \epsfbox {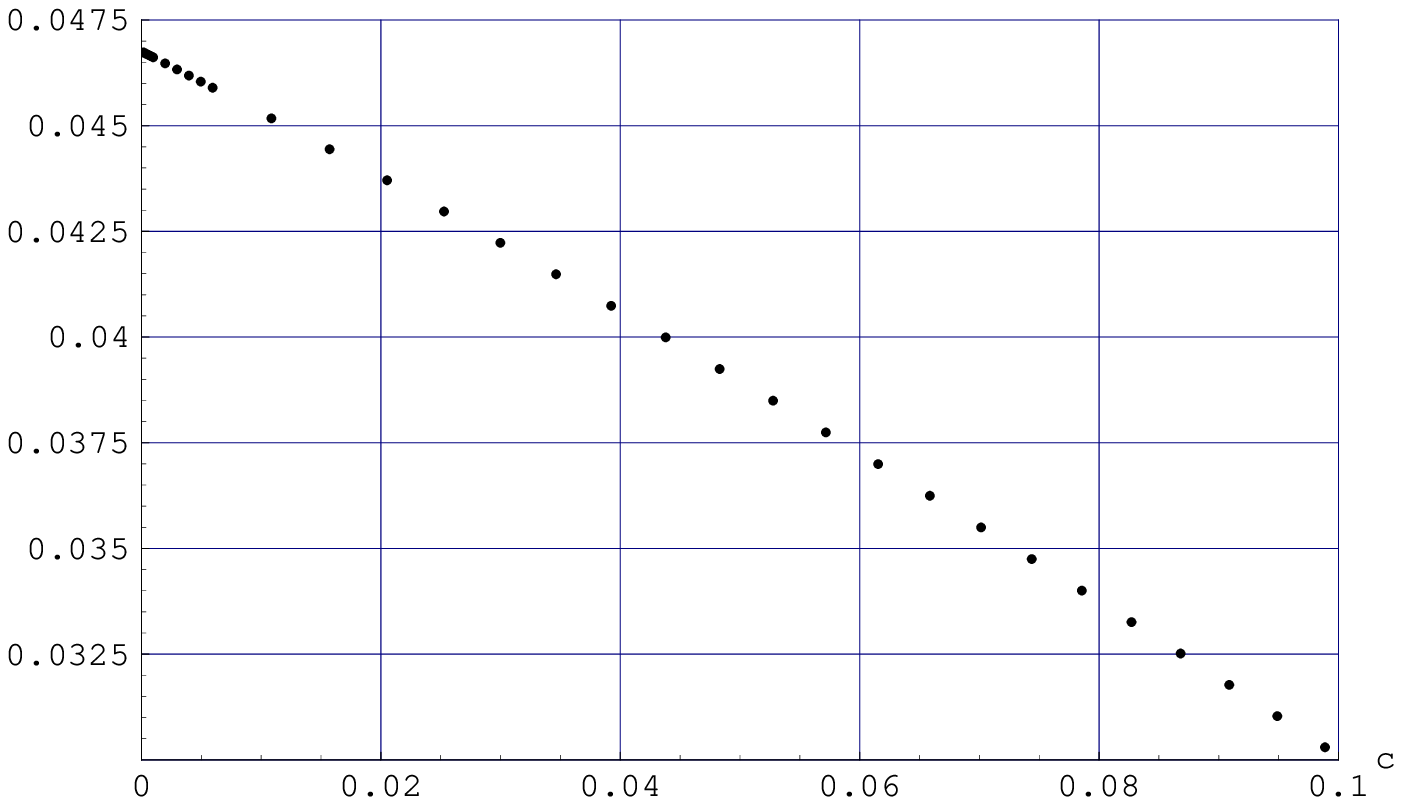}
}
}
\caption[a]{\it Behavior of $\gamma$ in the fine-monopole limit $c
\to 0$ for $\alpha=\frac{1}{2}$.}
\label{fmlgamma}
\end{figure}
\begin{figure}[htbp]
\centerline{
\parbox{9.0cm}{
$\scriptstyle \beta^2\gamma=\frac{\Lambda_7}{e^2 M_7^{10}}$ \\
\epsfxsize = 9.0cm \epsfbox {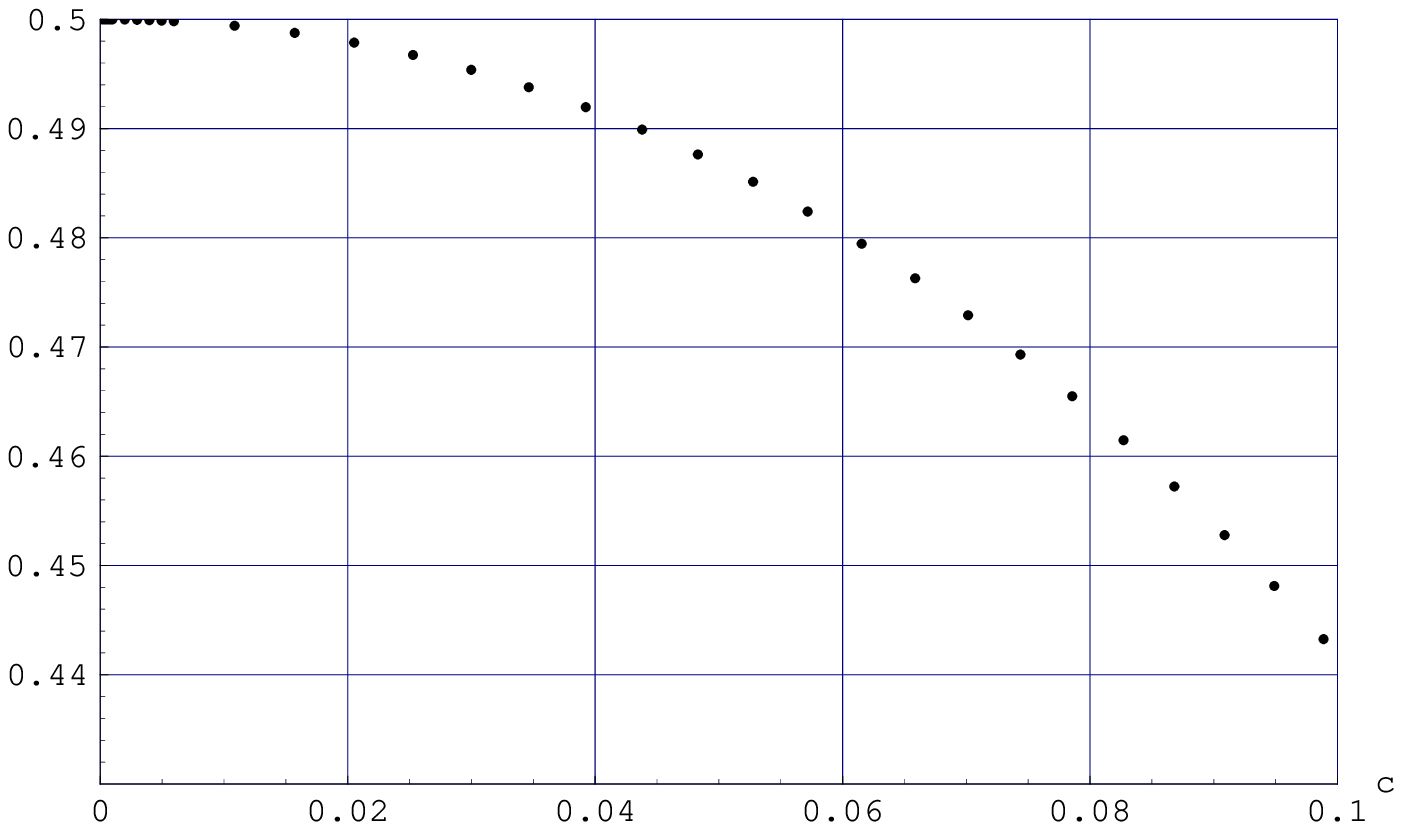}
}
}
\caption[a]{\it Behavior of $\beta^2 \gamma=\frac{\Lambda_7}{e^2 \,
 M_7^{10}}$ in the fine-monopole limit $c \to 0$ for
 $\alpha=\frac{1}{2}$.}
\label{fmlbgsqr}
\end{figure}
\begin{figure}[htbp]
\centerline{
\parbox{9.0cm}{
$\scriptstyle M_0$ \\
\epsfxsize = 9.0cm \epsfbox {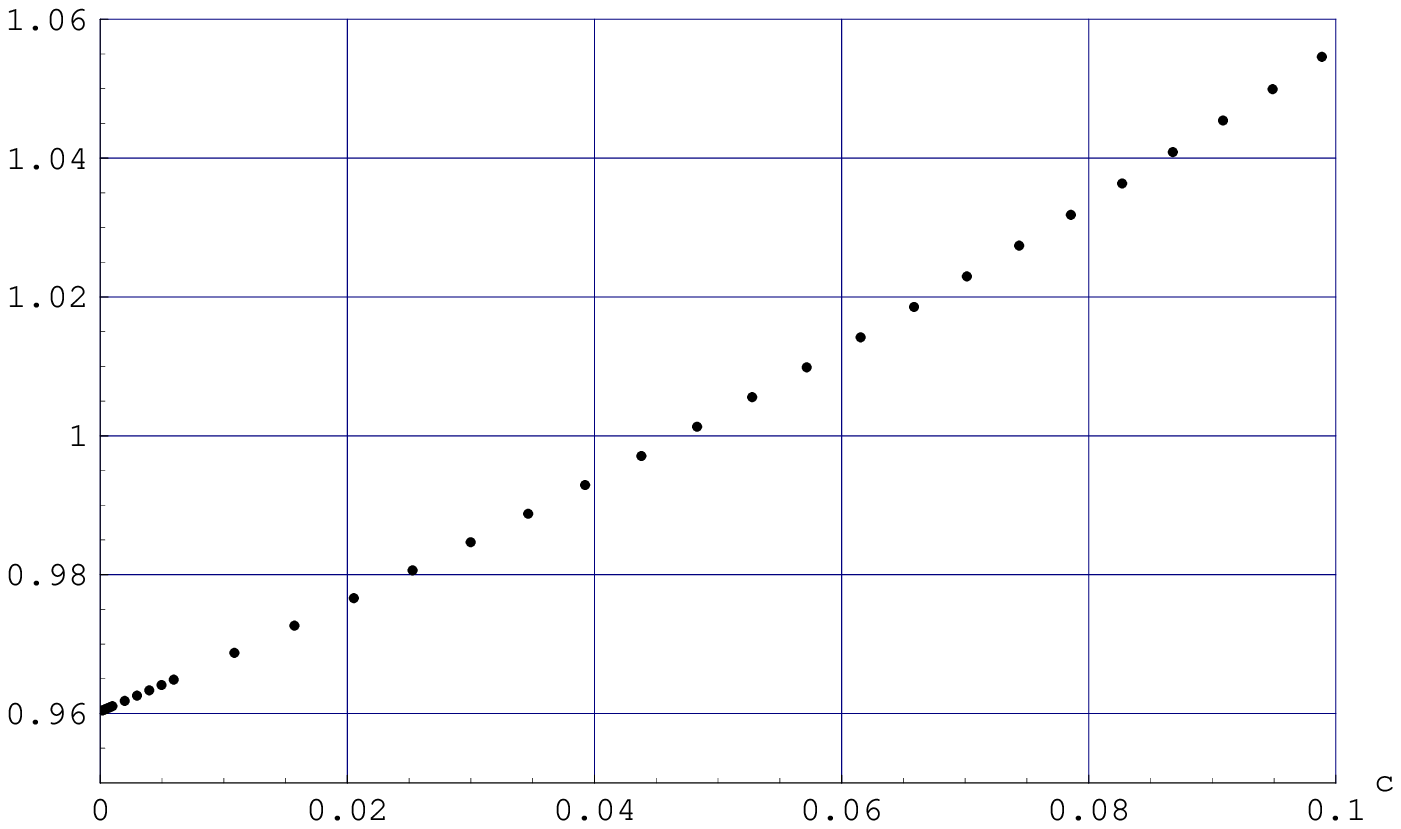}
}
}
\caption[a]{\it Behavior of $M_0$ in the fine-monopole limit $c \to
0$ for $\alpha=\frac{1}{2}$.}
\label{fmlM0}
\end{figure}

Note that $\beta^2 \gamma$ tends to $\frac{1}{2}$ in the
fine-monopole limit (see Fig.~\ref{fmlbgsqr}). Neglecting all orders
different from $\frac{1}{c}$  in the above expression for $M_P^2$ and
using (\ref{sizeofexdim}) one has
  \begin{equation} \label{Wmass}
    m_W=1.1 \cdot \, 10^{-5} \, \frac{\kappa^{5/2}}{\xi^{1/2}}
\mbox{GeV} \, ,
  \end{equation}
  where $r_0$ has been parametrized by $r_0=\frac{0.2 \, \mbox{\tiny
   mm}}{\xi}\approx 10^{12}\, \mbox{GeV}^{-1}  \cdot \frac{1}{\xi}$.
  \item Since Newtons law is established down to $0.2 \, \mbox{mm}$
\cite{submm} we simply need to have $\xi>1$.
  \item In order to have classical gravity applicable in the bulk we
require that curvature at infinity is negligible with respect to the
corresponding power of the fundamental scale. By looking at eqs. 
(\ref{curvinvinfinity1})-(\ref{curvinvinfinity4}) we see that we have
to impose 
  \begin{equation} 
    c^2 \, m_W^2 \ll M_7^2 \quad \mbox{and} \quad
\frac{m_W^2}{\mathcal{L}_0^2} \ll M_7^2 \, .
  \end{equation}
Using (\ref{EinsteininfI}) and the first of the above relations, the
second one can immediately be transformed into 
\begin{equation}
  \beta \, \gamma \, m_W^2 \ll M_7^2 \, .
\end{equation}
One then finds 
\begin{equation} \label{reqiii}
  2 \cdot 10^{-15} \frac{\xi}{\kappa} \ll 1 \quad \mbox{and} \quad
  4 \cdot 10^{-3} \left( \frac{\kappa^3}{\xi} \right)^{1/2} \ll 1\, ,
\end{equation}
where $\beta$ and $\gamma$ have again been replaced by their
limiting values (\ref{finemononum})  for $c \to 0$.  \item \label{4}
Classical gravity is applicable in the monopole core whenever the
curvature invariants $R^2$, $R_{A B} R^{A B}$, $R_{A B C D} R^{A B C
D}$ and $C_{A B C D} C_{A B C D}$ are small compared to the forth 
power of the fundamental gravity scale. Since these quantities are of
the order of the mass $m_W^4$  (see (\ref{curvatorigin})) we have
\begin{equation} 
\label{reqiv}
  m_W^4 \ll M_7^4 \, .
\end{equation} 
Using (\ref{fundgravscale}) and (\ref{Wmass}) one finds
\begin{equation}
  1.1 \cdot 10^{-8} \left( \frac{\kappa^3}{\xi} \right)^{1/2} \ll 1\, .
\end{equation}
\end{enumerate}
Finally, the fine-monopole limit requires 
\begin{equation} 
\label{thincondition}
  c=\frac{2}{r_0 \, m_W} \ll 1 \, .
\end{equation}
Again (\ref{Wmass}) implies
\begin{equation} \label{thinmonolimit}
  1.8 \cdot 10^{-7} \left(\frac{\xi^3}{\kappa^5}\right)^{1/2}\ll 1\, .
\end{equation}
It is now easy to see that for a wide range of parameter combinations
all these requirements on $\xi$ and $\kappa$ can simultaneously be
satisfied. One possible choice is $\xi=100$ and $\kappa=1$. This
shows that already in the case $\alpha=\frac{1}{2}$ there are
physical solutions corresponding to a fine-monopole in the sense of
eq.~(\ref{thincondition}) respecting all of the above requirements
(\ref{R1})-(\ref{R4}). Fig.~\ref{solFML} shows the fine-monopole 
solution corresponding to the lowest $c$-value in the sequence shown
in Figs.~\ref{fmlbeta} to \ref{fmlM0}.

\section{Conclusion}  \label{C}
We have demonstrated in this paper that it is possible to generalize
the idea of warped compactification on  a topological defect in a
higher dimensional spacetime to $n=3$ transverse dimensions by
considering a specific field theoretical model. Numerical solutions
were found for the case of a monopole realized as a 't Hooft-Polyakov
monopole. This generalization turned out to be non-trivial since (at
least) in the rotational invariant transverse  space setup
considered, Einstein's equations don't seem to admit strictly local
defect solutions. Even though the  transverse space still approaches
a constant curvature space at spatial infinity, it is not necessarily
an anti-de-Sitter space. Both signs of the bulk cosmological constant
are possible in order to localize gravity. We considered a fine 
monopole limit in the case $\alpha=1/2 \, (m_W=m_H)$ and verified
that the model proposed is not in conflict with  Newtons law, that it
leads to a possible solution of the hierarchy problem and that
classical gravity is applicable in the bulk and in the core of the
defect. Even though stability should be guaranteed by topology, small
perturbations around the monopole background should be considered as
well as quadratic corrections to the Einstein-Hilbert action.

{\it Acknowledgments:} We thank T. Gherghetta, H. Meyer, S.
Randjbar-Daemi,  P. Tinyakov, S. Wolf and K. Zuleta for helpful
discussions.  This work was supported by the FNRS grant 20-64859.01.

\appendix 

\section{Derivation of Fine-tuning relations} \label{ApA}
By taking linear combinations of Einstein equations
 (\ref{EinsteinP0}) to (\ref{EinsteinPtheta}) one can easily derive
 the following relations:
\begin{eqnarray}
  \frac{\left[M(r)^3 M'(r) \mathcal{L}(r)^2 \right]'}{M(r)^4
  \mathcal{L}(r)^2}&=& -\frac{\beta}{5} \left( 2 \gamma + \epsilon_0
  - \epsilon_\rho - 2 \epsilon_\theta \right) \, ,
  \label{Einsteincomb1}\\ 
  \frac{\left[M(r)^4 \mathcal{L}(r)
  \mathcal{L}'(r)\right]'}{M(r)^4
  \mathcal{L}^2(r)}-\frac{1}{\mathcal{L}(r)^2}&=& -\frac{\beta}{5}
  \left( 2 \gamma - 4 \epsilon_0 - \epsilon_\rho + 3 \epsilon_\theta
  \right) \, . 
  \label{Einsteincomb2}
\end{eqnarray}
By multiplying with $M(r)^4 \mathcal{L}(r)^2$, integrating from $0$
to $\infty$ and using the definition of  the brane tensions
(\ref{branetensions}) we obtain
\begin{eqnarray}
  M(r)^3 M'(r) \mathcal{L}(r)^2 \vert_{0}^{\infty}&=& 
-\frac{2\beta\gamma}{5} \int_{0}^{\infty} M(r)^4 \mathcal{L}(r)^2 dr 
+ \frac{\beta}{5} \left( \mu_{0}-\mu_{\rho}-2 \mu_{\theta}\right) \,
, \label{A1}\\ M(r)^4 \mathcal{L}(r) \mathcal{L}'(r)
\vert_{0}^{\infty}&=&\hspace{-2mm}\int_{0}^{\infty} \hspace{-2mm}
M(r)^4 dr -  \frac{2 \beta \gamma}{5} \int_0^{\infty} \hspace{-2mm}
M(r)^4 \mathcal{L}(r)^2 dr - \frac{\beta}{5}  \left(4 \mu_0 +
\mu_\rho - 3 \mu_\theta \right) \, . \label{A2} \end{eqnarray}
Using the boundary conditions for the metric functions
(\ref{BCMorigin}) and (\ref{asympto}), and taking the difference  of
the eqs.~(\ref{A1}) and (\ref{A2}) then establishes  the first part
of eq.~(\ref{FineTunRel1}). To prove the second part of
(\ref{FineTunRel1}) one starts right from the general expressions of
the stress-energy components $\epsilon_i$,
relations~(\ref{enmomelements1})-(\ref{enmomelements3}):
\begin{equation}
  \mu_0-\mu_{\theta}=\int_0^{\infty} dr M(r)^4 \mathcal{L}(r)^2
  \left( \epsilon_\theta - \epsilon_0 \right) = \int_0^{\infty} dr
  M(r)^4 \mathcal{L}(r)^2 \left\{ 
  \frac{K'(r)^2}{\mathcal{L}(r)^2}+
  \frac{\left[1-K(r)^2\right]^2}{\mathcal{L}(r)^4}
  +  \frac{J(r)^2 K(r)^2}{\mathcal{L}(r)^2} \right\} \, . 
\end{equation}
Multiplying the equation of motion for the gauge field (\ref{EqMovW})
by $K(r)$ and substituting  the $J(r)^2 K(r)^2$ term gives 
\begin{eqnarray}
  \mu_0-\mu_\theta=\int_0^{\infty} dr M(r)^4 \left\{ K'(r)^2+\frac{1-
     K(r)^2}{\mathcal{L}(r)^2} + \frac{\left[ M(r)^4 K'(r) \right]'
     K(r)}{M(r)^4} \right\}  \, .
\end{eqnarray}
Integration by parts in the last term of the above equation leads to
\begin{eqnarray}
   \mu_0-\mu_\theta=\int_0^{\infty} dr M(r)^4
   \left[\frac{1-K(r)^2}{\mathcal{L}(r)^2}\right]+  K(r) M(r)^4 K'(r)
   \vert_0^{\infty} \, .
\end{eqnarray} 
which together with the behavior of the gauge field at the origin 
$K'(0) =0$ (see (\ref{asy_origin_K})) finishes the proof of relation 
(\ref{FineTunRel1}). 

The proof of relation (\ref{FineTunRel2}) is simply obtained by
rewriting (\ref{A1}) with vanishing left hand side.

To establish (\ref{FineTunRel3}) we start directly from the
definitions of the stress-energy tensor components,
relations~(\ref{enmomelements1})-(\ref{enmomelements3}):
\begin{equation} \label{FTR3int}
  \mu_0+\mu_{\rho}+2\mu_{\theta}= \int_{0}^{\infty} dr M(r)^4
    \mathcal{L}(r)^2  \left[ J'(r)^2+\frac{2 \, J(r)^2
    K(r)^2}{\mathcal{L}(r)^2}+\alpha \left( J(r)^2-1\right)^2 \right]
    \, . 
\end{equation}
Collecting derivatives in the equation of motion for the scalar field
(\ref{EqMovphi}) and multiplying by $J(r)$ gives
\begin{equation}
  \frac{2 \, J(r)^2 K(r)^2}{\mathcal{L}(r)^2} =  \frac{\left[ M(r)^4
    \mathcal{L}(r)^2 J'(r) \right]'}{M(r)^4 \mathcal{L}(r)^2} J(r) 
    -\alpha J(r)^2 \left( J(r)^2-1\right) \, .
\end{equation}
Eliminating now the second term in the equation (\ref{FTR3int}) leads 
to 
\begin{equation}
  \mu_0+\mu_{\rho}+2\mu_{\theta}= \int_{0}^{\infty} dr M(r)^4
   \mathcal{L}(r)^2 \left[ J'(r)^2+\frac{\left[ M(r)^4
   \mathcal{L}(r)^2 J'(r) \right]'}{M(r)^4 \mathcal{L}(r)^2} J(r) +
   \alpha \left( 1- J(r)^2 \right) \right] \, .
\end{equation}
If we now expand and integrate the second term by parts we are left
with
\begin{equation} 
 \mu_0+\mu_{\rho}+2\mu_{\theta}= \alpha \int_{0}^{\infty} dr \left(
 1- J(r)^2 \right) M(r)^4 \mathcal{L}(r)^2 + M(r)^4 \mathcal{L}(r)^2
 J(r) J'(r) \vert_0^{\infty} \, ,
\end{equation}
which reduces to (\ref{FineTunRel3}) when the boundary conditions for
the metric  (\ref{BCMorigin}) and (\ref{asympto}) are used.

\section{Numerics} \label{ApB}

As already pointed out, the numerical problem encountered is to find
those solutions to the system of differential  equations
(\ref{EinsteinP0})-(\ref{EqMovW}) and boundary conditions for which
the integral  defining the $4$-dimensional Planck-scale
(\ref{PlanckMass}) is finite. This is a two point  boundary value
problem on the interval $r=[0,\infty)$ depending on three independent
parameters $(\alpha,\beta,\gamma)$. Independently of the numerical
method employed, the system of equations 
(\ref{EinsteinP0})-(\ref{EqMovW}) was rewritten in a different way in
order for the integration to be as stable as possible. By introducing
the derivatives of the unknown functions $J(r)$, $K(r)$, $M(r)$ and
$\mathcal{L}(r)$ as new dependent variables one obtains a system of
ordinary first order equations. In the case of $M(r)$ it has proven
to be convenient to define $y_7=M'(r)/M(r)$ as a new unknown function
(rather than $M'(r)$) since the boundary condition for $y_7$ at
infinity then simply reads
\begin{equation}
  \lim_{r\to \infty} y_7=-\frac{c}{2} \, .
\end{equation}
With the following definitions we give the form of the 
equations which is at the base of several numerical methods employed:
\begin{equation} 
\label{numsys}
  \begin{array}{lclrcl}
    y_1(r) &=& J(r)\, ,& \quad y'_1 &=& y_5 \, , \\  
    y_2(r) &=& K(r)\, ,&  y'_2 &=& y_6 \, , \\  
    y_3(r) &=& M(r)\, ,&  y'_3 &=& y_3 \, y_7 \, , \\  
    y_4(r) &=& \mathcal{L}(r)\, ,& y'_4 &=& y_8 \, , \\  
    y_5(r) &=& J'(r)\, ,& y'_5 &=& -2 \left( 2 \,
	       y_7+\frac{y_8}{y_4}\right) y_5+2\, \frac{y_1
	       y_2^2}{y_4^2}+ \alpha \, y_1 \left( y_1^2-1 \right) \,
	       , \\  
    y_6(r) &=& K'(r)\, ,& y'_6 &=& -\frac{y_2 \left( 1- y_2^2
		\right)}{y_4^2} - 4 y_6 y_7 + y_1^2 y_2 \, , \\  
    y_7(r) &=& M'(r)/M(r)\, ,& y'_7 &=& -4 y_7^2 - 2 y_7
	       \frac{y_8}{y_4} - \frac{\beta}{5} \left( 2 \gamma +
	       \epsilon_0 - \epsilon_{\rho}-2 \epsilon_{\theta}
	       \right) \, , \\  
    y_8(r) &=& \mathcal{L}'(r)\, ,& y'_8 &=& -\frac{y_8^2}{y_4}-4 y_7
		y_8 + \frac{1}{y_4}- \frac{\beta y_4}{5} \left( 2
		\gamma -4 \epsilon_0 - \epsilon_{\rho}+3
		\epsilon_{\theta} \right) \, ,
  \end{array}
\end{equation}
with 
\begin{eqnarray} 
\label{epscomb}
  \epsilon_0 - \epsilon_{\rho}-2 \epsilon_{\theta} &=& 
       \frac{\alpha}{2} \left( y_1^2 - 1 \right)^2 
      -\frac{\left(1-y_2^2\right)^2}{y_4^4}-2 \, \frac{y_6^2}{y_4^2}
      \, , \\
  -4 \epsilon_0 - \epsilon_{\rho}+3 \epsilon_{\theta} &=&
       \frac{\alpha}{2} \left( y_1^2 - 1 \right)^2 
   + 4 \, \frac{\left(1-y_2^2\right)^2}{y_4^4}
   + 5 \, \frac{y_1^2 \, y_2^2}{y_4^2} + 3 \, \frac{y_6^2}{y_4^2} \, . 
\end{eqnarray}
This is an autonomous ordinary system of coupled differential
equations depending on the parameters $(\alpha,\,\beta,\,\gamma)$.
The boundary conditions are 
\begin{equation} 
  \begin{array}{rcl} y_1(0)&=&0 \, , \\ 
  \lim\limits_{r\to\infty} y_1(r) &=& 1 \, , 
  \end{array} \quad
  \begin{array}{rcl} y_2(0)&=&1 \, , \\ 
  \lim\limits_{r\to\infty} y_2(r) &=& 0 \, , 
  \end{array} \quad
  \begin{array}{rcl} y_3(0)&=&1 \, , \\ y_7(0) &=& 0 \, , 
  \end{array}\quad
  \begin{array}{rcl} y_4(0)&=&0 \, , \\ y_8(0) &=& 1 \, .  
  \end{array}\quad
\end{equation}

In order to find solutions with the desired metric asymptotics at
infinity it is useful to define either one (or more) of the
parameters $(\alpha,\,\beta,\,\gamma)$ or the constants $J'(0)$ and
$K''(0)$ as additional dependent variables,  e.g. $y_9(r)=\alpha$
with $y'_9(r)=0$, see \cite{numrec}. Before discussing the different
methods that were used we give some common numerical problems
encountered.
\begin{itemize} 
\item Technically it is impossible to integrate to infinity. The
possibility of compactifying the independent variable $r$ was not
believed to simplify the numerics. Therefore the integration has to
be stopped at some upper  value of $r=r_{\mbox{\tiny max}}$. For most
of the solutions this was about $20 \sim 30$.
\item Forward integration with arbitrary but fixed values of
$(\alpha,\,\beta,\,\gamma,\, J'(0)\, ,K''(0))$ turned out to be very
unstable. This means that even before some integration routine (e.g.
the Runge-Kutta method \cite{numrec}) reached $r_{\mbox{\tiny max}}$,
the values of some $y_i$ went out of range which was due to the
presence of terms  $\frac{1}{\left(\mathcal{L}\right)^n}$ or
quadratic and cubic (positive coefficient) terms in $y_i$.
\item Some right hand sides of (\ref{numsys}) contain terms singular
at the origin such that their sums remain regular. Starting the
integration at $r=0$ is therefore impossible. To overcome this
problem the solution in terms of the power series
(\ref{asy_origin_M})-(\ref{asy_origin_K}) was used within the
interval $\left[ 0, \,  \epsilon \right]$ (for $\epsilon=0.01$).
\item The solutions are extremely sensitive to initial conditions
which made it unavoidable to pass from single precision to double
precision (from about $7$ to about $15$ significant digits). However
this didn't completely solve the  problem. Even giving initial
conditions (corresponding to a gravity localizing solution) at the
origin to machine  precision is in general not sufficient to obtain
satisfactory precision at $r_{\mbox{\tiny max}}$ in a single 
Runge-Kutta forward integration step from $\epsilon$ to 
$r_{\mbox{\tiny max}}$.
\end{itemize}

The method used for exploiting the parameter space
$(\alpha,\,\beta,\,\gamma)$ for gravity localizing solutions  was a
generalized version of the shooting method called the multiple
shooting method \cite{numrec}, \cite{acton},  \cite{keller},
\cite{BulirschStoer}. In the shooting method a boundary value problem
is solved by combining a root-finding method (e.g. Newton's method
\cite{numrec}) with forward integration. In order to start the
integration at one boundary, the root finding  routine specifies
particular values for the so-called shooting parameters and compares
the results of the integration with the boundary conditions at the
other boundary. This method obviously fails whenever the ``initial
guess'' for the  shooting parameters is too far from a solution such
that the forward integration does not reach the second boundary. For
this reason the multiple shooting method was used in which the
interval  $\left[ \epsilon \, , r_{\mbox{\tiny max}} \right]$ was
divided into an variable number of sub-intervals in  each of which the
shooting method was applied. This of course drastically increased the
number of shooting  parameters and as a result the amount of
computing time. However, it resolved two problems:
\begin{itemize}
\item Since the shooting parameters were specified in all
sub-intervals the precision of the parameter values  at the origin was
no longer crucial for obtaining high precision solutions.
\item Out of range errors can be avoided by augmenting the number of
shooting intervals (at the cost of  increasing computing time).
\end{itemize}

Despite all these advantages of multiple shooting, a first
combination of parameter values  $(\alpha,\,\beta,\,\gamma,\, J'(0)\,
,K''(0))$ leading to gravity localization could not be found by this
method, since convergence depends  strongly on how close initial
shooting parameters are to a real solution. This first solution was
found by backward integration combined with the simplex method for
finding zeros of one real function  of several real variables. For a
discussion of the simplex method see e.g. \cite{numrec}. This
solution,  corresponds to the parameter values $(\alpha=1.429965428 ,
\, \beta=3.276535576 , \, \gamma=0.025269415) $. 

Once this solution was known, it was straightforward to investigate
with the multiple shooting method  which subset of the $(\alpha,\,
\beta,\, \gamma)$-space leads to the desired metric asymptotics. We
used a  known solution $(\alpha_1, \, \beta_1, \,\gamma_1)$ to obtain
starting values for the shooting parameters of a closeby other
solution $(\alpha_1+\delta\alpha,\,\beta_1 + \delta\beta,\,\gamma_1 +
\delta\gamma)$. This lead in general to rapid convergence of the
Newton-method. Nevertheless, $\delta\alpha, \, \delta \beta,\,\delta
\gamma$ still had to be small. By this simple but time-consuming
operation the fine-tuning surface  in  parameter-space, presented in 
section \ref{NFTS} and shown in Fig.~\ref{FTS} was found.


\begin{thebibliography}{99}

\bibitem{RS2} V.~A.~Rubakov and M.~E.~Shaposhnikov, 
Phys.\ Lett.\  {\bf B125} (1983) 136.

\bibitem{akama} K.~Akama, in {\it Proceedings of the Symposium on
Gauge Theory and  Gravitation}, Nara, Japan, eds. K.~Kikkawa,
N.~Nakanishi and H.~Nariai  (Springer-Verlag, 1983),
[hep-th/0001113].

\bibitem{KK} {\it Modern Kaluza-Klein Theories},  
eds.T.~Appelquist, A.~Chodos and P.~G.~Freund, 
(Addison-Wesley, 1987).

\bibitem{DS} 
G.~Dvali, M.~Shifman, Phys.\ Lett.\ {\bf B396} (1997) 64-69;  
[Erratum-ibid.
{\bf B407} (1997) 452]

\bibitem{Arkani-Hamed:1998rs}
N.~Arkani-Hamed, S.~Dimopoulos and G.~R.~Dvali,
Phys.\ Lett.\ B {\bf 429} (1998) 263
[arXiv:hep-ph/9803315].

\bibitem{rusu2} L.~Randall and R.~Sundrum, 
Phys.\ Rev.\ Lett.\  {\bf 83} (1999) 4690[hep-th/9906064].

\bibitem{rusu1} L.~Randall and R.~Sundrum,
 Phys.\ Rev.\ Lett.\  {\bf 83} (1999)3370 [hep-ph/9905221].

\bibitem{Rubakov:1983bz}
V.~A.~Rubakov and M.~E.~Shaposhnikov,
Phys.\ Lett.\ B {\bf 125} (1983) 139.

\bibitem{Randjbar-Daemi:1985wg}
S.~Randjbar-Daemi and C.~Wetterich,
Phys.\ Lett.\ B {\bf 166} (1986) 65.

\bibitem{Arkani-Hamed:2000eg}
N.~Arkani-Hamed, S.~Dimopoulos, N.~Kaloper and R.~Sundrum,
Phys.\ Lett.\ B {\bf 480} (2000) 193
[arXiv:hep-th/0001197].

\bibitem{Dvali:2002pe}
G.~Dvali, G.~Gabadadze and M.~Shifman,
arXiv:hep-th/0202174.

\bibitem{branes} J.~Polchinski,
Phys.\ Rev.\ Lett.\  {\bf 75} (1995) 4724
[hep-th/9510017].

\bibitem{DeWolfe:1999cp}
O.~DeWolfe, D.~Z.~Freedman, S.~S.~Gubser and A.~Karch,
Phys.\ Rev.\ D {\bf 62} (2000) 046008
[arXiv:hep-th/9909134].

\bibitem{ck} A.~G.~Cohen and D.~B.~Kaplan, 
Phys.\ Lett.\  {\bf B470}
(1999) 52 [hep-th/9910132].

\bibitem{cp} A.~Chodos and E.~Poppitz, 
Phys.\ Lett.\  {\bf B471} (1999) 119 [hep-th/9909199].

\bibitem{ov}
I.~Olasagasti and A.~Vilenkin,
Phys.\ Rev.\ D {\bf 62} (2000) 044014
[arXiv:hep-th/0003300].

\bibitem{rg} R.~Gregory, 
Phys.\ Rev.\ Lett.\  {\bf 84} (2000) 2564
[hep-th/9911015].

\bibitem{gs}
T.~Gherghetta and M.~E.~Shaposhnikov,
Phys.\ Rev.\ Lett.\  {\bf 85} (2000) 240
[arXiv:hep-th/0004014].

\bibitem{Harvey}
M.~Giovannini, H.~Meyer and M.~E.~Shaposhnikov,
Nucl.\ Phys.\ B {\bf 619} (2001) 615
[arXiv:hep-th/0104118].

\bibitem{KP} P.~Tinyakov, K.~Zuleta, 
Phys.\ Rev.\ D {\bf 64} (2001) 025022.

\bibitem{Randjbar-Daemi:1982hi}
S.~Randjbar-Daemi, A.~Salam and J.~Strathdee,
Nucl.\ Phys.\ B {\bf 214} (1983) 491.

\bibitem{Randjbar-Daemi:1983qa} S.~Randjbar-Daemi, A.~Salam and
J.~Strathdee, Phys.\ Lett.\ B {\bf 132} (1983) 56.

\bibitem{GRS} T.~Gherghetta, E.~Roessl, M.~E.~Shaposhnikov, 
Phys.\ Lett.\  {\bf B491} (2000) 353.

\bibitem{dvali}
G.~R.~Dvali,
arXiv:hep-th/0004057.

\bibitem{bc} 
K.~Benson, I.~Cho  Phys.\ Rev.\ {\bf D64} (2001) 065026

\bibitem{Randjbar-Daemi:2000ft}
S.~Randjbar-Daemi and M.~E.~Shaposhnikov,
Phys.\ Lett.\ B {\bf 491} (2000) 329
[arXiv:hep-th/0008087].

\bibitem{Randjbar-Daemi:2000cr}
S.~Randjbar-Daemi and M.~E.~Shaposhnikov,
Phys.\ Lett.\ B {\bf 492} (2000) 361
[arXiv:hep-th/0008079].

\bibitem{tHooft} G.~'t Hooft, Nucl.\ Phys.\ {\bf B79} (1974) 276.

\bibitem{Polyakov} A.M.~Polyakov,  JETP\ Lett.\ {\bf 20}, (1974) 194

\bibitem{nwp} P.~van Nieuwenhuizen, D.~Wilkinson and M.~J.~Perry,
Phys.\ Rev.\ {\bf D13} (1976) 778.

\bibitem{br} F.~A.~Bais and R.~J.~Russell 
Phys.\ Rev.\ {\bf D11} (1975) 2692.

\bibitem{lnw} K.~Lee, V.P.~Nair and E.J.~Weinberg, Phys.\ Rev.\ {\bf
D45} (1992) 2751.

\bibitem{ortiz} M.~E.~Ortiz, 
Phys.\ Rev.\ {\bf D45} (1992) R2586.

\bibitem{bfm} P.~Breitenlohner, P.~Forgács and D.~G.~Maison,
Nucl.\ Phys.\  {\bf B383} (1992) 357.

\bibitem{wigner} C.~W.~Misner, K.~S.~ Thorne and J.~A.~Wheeler,
``Gravitation'' , Freeman and Company, San Francisco (1973)

\bibitem{GG} M.~Georgi, S.~L.~Glashow,  Phys.\ Rev.\ Lett.\ {\bf 28}
(1972) 1494.

\bibitem{Vilenkin} A.~Vilenkin, E.~P.~S.~Shellard, ``Cosmic Strings
and Other Topological Defects'', Cambridge Monographs on Mathematical
Physics (1994)

\bibitem{Raj} R.~Rajaraman, ``Solitons and Instantons, An
Introduction to Solitons and Instantons in Quantum Field Theory'', 
North Holland (1984)

\bibitem{PraSom} M.K.~Prasad, C.H.~Sommerfield, Phys.\ Rev.\ Lett.
{\bf 35} (1975) 760

\bibitem{submm} C.~D.~Hoyle et al., Phys. Rev. Lett. {\bf 86}, 1418
(2001).

\bibitem{numrec} W.~H.~Press, S.~A.~Teukolsky, W.~T.~Vetterling,
B.~P.~Flannery, ``Numerical Recipes in C, The Art of  Scientific
Computing'', Cambridge University Press, Second Edition (1992)

\bibitem{acton} F.~S.~Acton, ``Numerical Methods That Work'',
Washington: Mathematical Association of America, corrected edition
1990.  

\bibitem{keller} H.~B.~Keller, ``Numerical Methods For Two-Point
Boundary-Value Problems'', Waltham, MA: Blaisdell, 1968

\bibitem{BulirschStoer} J.~Stoer, R.~Bulirsch, ``Introduction to
Numerical Analysis'' New York: Springer-Verlag, 1980

\end{thebibliography}
\end{document}